\documentclass[twocolumn]{aastex7}

\usepackage{booktabs}

\def\degr{^{\circ}}

\def\arcsec{^{\prime\prime}}

\def\bf{\textbf}
\def\tt{\texttt}

\begin{document}

\title{X-ray Polarization and Spectral Variations in an Extreme High-Synchrotron-Peaked Blazar 1ES 1101--232}

\correspondingauthor{Jin Zhang}
\email{j.zhang@bit.edu.cn}

\author{Xin-Ke Hu}
\affiliation{School of Physics, Beijing Institute of Technology, Beijing 100081, People's Republic of China}
\email{xk.hu@bit.edu.cn}

\author[orcid=0000-0003-3554-2996]{Jin Zhang\dag}
\affiliation{School of Physics, Beijing Institute of Technology, Beijing 100081, People's Republic of China}
\email{j.zhang@bit.edu.cn}

\author[orcid=0000-0002-0105-5826]{Fei Xie}
\affiliation{Guangxi Key Laboratory for Relativistic Astrophysics, School of Physical Science and Technology, Guangxi University, Nanning 530004, People's Republic of China}
\affiliation{INAF Istituto di Astrofisica e Planetologia Spaziali, Via del Fosso del Cavaliere 100, 00133 Roma, Italy}
\email{xief@gxu.edu.cn}

\author[orcid=0000-0001-8411-8011]{Xiang-Gao Wang}
\affiliation{Guangxi Key Laboratory for Relativistic Astrophysics, School of Physical Science and Technology, Guangxi University, Nanning 530004, People's Republic of China}
\affiliation{GXU-NAOC Center for Astrophysics and Space Sciences, Nanning 530004, People's Republic of China}
\email{wangxg@gxu.edu.cn}

\begin{abstract}

We present the first X-ray polarimetric observation of the extreme high-synchrotron-peaked blazar 1ES 1101--232, conducted by the Imaging X-ray Polarimetry Explorer (IXPE). The data analysis incorporates simultaneous and quasi-simultaneous observations from Swift-XRT and NuSTAR. Our results reveal a significant detection of X-ray polarization in the 2--6 keV band at a confidence level (CL) of 6.6$\sigma$, with a polarization degree of $\Pi_{\rm X}=17.9\%\pm2.7\%$ and an electric vector position angle (EVPA) of $\psi_{\rm X}=10\degr.0\pm4\degr.4$. An even higher polarization degree of $\Pi_{\rm X}=38.9\%\pm9.1\%$ with an EVPA of $\psi_{\rm X}=13\degr.9\pm6\degr.7$ is observed within a narrower time interval, at a CL of 4.3$\sigma$. During the IXPE observational campaign, the X-ray spectrum of 1ES 1101--232 exhibits a clear soft-to-hard spectral evolution in the 0.3--10 keV band, although no significant flux variability is detected. Additionally, a clockwise hysteresis loop is identified in the flux--photon index plane. These findings collectively indicate that the X-ray emission from 1ES 1101--232 originates in a region characterized by a well-ordered magnetic field through synchrotron radiation.

\end{abstract}


\section{Introduction}\label{sec:intr}

Blazars are an extremely active subclass of active galactic nuclei (AGNs), characterized by highly relativistic plasma jets that are nearly aligned with observers' line of sight \citep[e.g.,][]{1995PASP..107..803U,2019ARA&A..57..467B,2019NewAR..8701541H}. Due to the effects of relativistic beaming, such alignment causes the non-thermal radiation from the jets to dominate the broadband emission spectrum, which spans from radio waves to very-high-energy (VHE) $\gamma$-rays \citep{1999ApJ...521..493L}. Therefore, their jets, which are observable across the entire electromagnetic spectrum, serve as excellent laboratories for investigating the acceleration, cooling, and interaction processes of relativistic particles in some of the Universe's most extreme environments.

The broadband spectral energy distributions (SEDs) of blazars usually exhibit a double-hump structure. It is widely accepted that the low-energy hump (peaking in the infrared--X-ray bands) is produced by synchrotron radiation from relativistic electrons. In contrast, the origin of the high-energy hump (peaking in the MeV--TeV $\gamma$-ray bands) remains uncertain. Models explaining the high-energy emission can be mainly classified into two scenarios: leptonic and hadronic. Leptonic models attribute the high-energy emission to inverse-Compton scattering, primarily the synchrotron self-Compton (SSC) process \citep[e.g.,][]{1992ApJ...397L...5M,1996MNRAS.280...67G,1996A&AS..120C.503G,2009ApJ...704...38S,2012ApJ...752..157Z,2014MNRAS.439.2933Y} or the external Compton process \citep[e.g.,][]{1993ApJ...416..458D,1994ApJ...421..153S,2000ApJ...545..107B,2009MNRAS.397..985G,2014ApJ...788..104Z,2015ApJ...807...51Z}. Hadronic models explain this emission through proton synchrotron radiation \citep[e.g.,][]{2000NewA....5..377A,2013ApJ...768...54B} or photo-pion production processes \citep[e.g.,][]{1993A&A...269...67M,2022ApJ...925L..19C}. Based on the peak frequency of the synchrotron component ($\nu_{\rm s,p}$), blazars are commonly categorized into three types \citep[e.g.,][]{2010ApJ...710.1271A,2016ApJS..226...20F}: low-synchrotron-peaked blazars (LSPs; $\nu_{\rm s,p}<10^{14}~{\rm Hz}$, peaking in the infrared bands), intermediate-synchrotron-peaked blazars (ISPs; $10^{14}~{\rm Hz}<\nu_{\rm s,p}<10^{15}~{\rm Hz}$, peaking in the optical--ultraviolet bands), and high-synchrotron-peaked blazars (HSPs; $\nu_{\rm s,p}>10^{15}~{\rm Hz}$, peaking in the ultraviolet--X-ray bands).

Polarization measurements, which provide information regarding the geometry and structure of magnetic fields, as well as the composition and turbulence in the jet, serve as a vital probe for testing models of particle acceleration and radiation. The polarization degree ($\Pi$), i.e., the fraction of polarized radiation, provides a measure of the degree of order in the magnetic field at the emission site \citep{2024A&A...689A.119K}. The polarization angle ($\psi$), that is, the electric vector position angle (EVPA), which defines the position angle of the linearly polarized emission, reveals the orientation of the mean magnetic field under the assumption that relativistic electrons are distributed isotropically in the co-moving frame \citep{2024A&A...689A.119K}. Simultaneous multiwavelength polarization observations can play a crucial role in determining whether the emission at different wavelengths is co-located. Polarization measurements in the past decades have been limited to wavelengths longer than X-ray energies. The Imaging X-ray Polarimetry Explorer \citep[IXPE;][]{2022JATIS...8b6002W}, successfully launched on 2021 December 9, is the first dedicated X-ray polarimetry mission, offering an unprecedented opportunity to measure the X-ray polarization properties of HSPs. In the past three years, IXPE has observed over ten blazars, with significant X-ray polarization detected in all observed HSPs. Notable examples include Mrk 421 \citep{2022ApJ...938L...7D}, Mrk 501 \citep{2022Natur.611..677L,2024ApJ...970L..22H}, PG 1553+113 \citep{2023ApJ...953L..28M}, 1ES 0229+200 \citep{2023ApJ...959...61E}, PKS 2155--304 \citep{2024ApJ...963L..41H,2024A&A...689A.119K}, and H 1426+428 \citep{2025ApJ...986..182H}. These observations support a model in which shock-accelerated, energy-stratified electron populations are responsible for the synchrotron X-ray emission in the jets of HSPs.

1ES 1101--232 is a TeV-emitting HSP, located at $z=0.186$ \citep{1989ApJ...345..140R}, in the Southern Hemisphere. Its TeV emission was first detected by the High Energy Stereoscopic System of atmospheric Cherenkov telescopes \citep{2007A&A...470..475A}. Additionally, 1ES 1101--232 belongs to a rare subclass of blazars, namely extreme high-synchrotron-peaked blazar (EHSP, $\nu_{\rm s,p}>10^{17}~{\rm Hz}$, peaking in the X-ray band). It is the third EHSP observed by IXPE following 1ES 0229+200 \citep{2023ApJ...959...61E} and H 1426+428 \citep{2025ApJ...986..182H}. BeppoSAX observations of 1ES 1101--232 demonstrated that the flux declined from $\sim3.8\times10^{-11}~{\rm erg~cm^{-2}~s^{-1}}$ in 1997 to $\sim2.5\times10^{-11}~{\rm erg~cm^{-2}~s^{-1}}$) in 1998 within the 2--10 keV band, and this was accompanied by a hard-to-soft spectral variation \citep{2000A&A...357..429W}. An observation conducted by Suzaku in 2006 indicated that 1ES 1101--232 was in a quiescent state, with an average 2--10 keV flux of $\sim1.68\times10^{-11}~{\rm erg~cm^{-2}~s^{-1}}$, exhibiting the lowest X-ray flux ever detected \citep{2008ApJ...682..775R}. Its X-ray spectrum was well fitted by a broken power-law or log-parabola model, with curvature extending up to the highest hard X-ray data point ($\sim30$ keV). A previous NuSTAR observation showed evidence for intraday variability in the 3--79 keV X-ray emission of 1ES 1101--232, with an unabsorbed flux of $\sim2.94\times10^{-11}~{\rm erg~cm^{-2}~s^{-1}}$ \citep{2018ApJ...859...49P}. As an EHSP, the synchrotron peak of 1ES 1101--232 might be located near or within the 2--8 keV band, and its flux, even in a quiescent state, reaches $10^{-11}~{\rm erg~cm^{-2}~s^{-1}}$ in the 2--10 keV band. Such characteristics make 1ES 1101--232 a valuable target for IXPE polarimetry.

In this paper, we report the results from the first IXPE observation of 1ES 1101--232, as well as from simultaneous and quasi-simultaneous X-ray observations performed by the Swift-XRT \citep{2004ApJ...611.1005G,2005SSRv..120..165B} and NuSTAR \citep{2013ApJ...770..103H}. In Section \ref{sec:obs&anal}, we introduce the X-ray observations used in this work and present the details of data analysis. Section \ref{sec:pol} reports the results derived from the IXPE data analysis. The spectral variation of X-rays from 1ES 1101--232 is studied in Section \ref{sec:spec}. A discussion of the key findings and conclusions are provided in Section \ref{sec:disc&conc}. Unless otherwise stated, all uncertainties quoted in the text and error bars displayed in the figures correspond to 1$\sigma$ (68.28\%) confidence level (CL) throughout the paper.

\section{Observations and Data Analysis}\label{sec:obs&anal}
\subsection{IXPE}\label{subsec:ixpe}

As part of a multiwavelength campaign, IXPE observed 1ES 1101--232 from 19:43 UTC on 2024 November 28 to 22:51 UTC on 2024 December 2, achieving a net exposure of $\sim191$ ks (OBSID: 03006901; PI: F. Tavecchio). More details of the observation information are presented in Table \ref{tab:obs}. In this work, we present a comprehensive analysis of the IXPE data for 1ES 1101--232 and interpret its X-ray polarization properties.

First, a World Coordinate System (WCS) correction was applied to the Level 2 event files to account for the detector pointing misalignment. As suggested in \citet{2023AJ....165..143D}, for an intermediate source like 1ES 1101--232, rejecting the IXPE instrumental background events is recommended and effective. Thus, we removed the instrumental background events from the Level 2 event files for all three detector units (DUs) using the \tt{filter\_background.py} script\footnote{\url{https://heasarc.gsfc.nasa.gov/docs/ixpe/analysis/contributed.html}}, which requires the corresponding Level 1 event files as input. Additionally, good time intervals were filtered to reduce particle events associated with solar activity \citep{2024ApJ...962...92X,2024ApJ...967L..38F,2024ApJ...974L...1K,2025MNRAS.540.3242D}. Given that 1ES 1101--232 is a point-like source within the IXPE angular resolution of $30\arcsec$ and its X-ray emission becomes faint beyond $45\arcsec$, we defined the source region as a circle with radius of $45\arcsec$ centered on the coordinates (R.A. = $165\degr.907$, decl. = $-23\degr.492$) provided by the Wide-field Infrared Survey Explorer \citep[WISE;][]{2010AJ....140.1868W}. Similarly, the background region was chosen as an annulus with inner and outer radii of $120\arcsec$ and $270\arcsec$, respectively, concentric with the source region. We used the \tt{xpselect} task within the software \tt{ixpeobssim} \citep[v.31.0.3;][]{2022SoftX..1901194B} to extract the source and background events from the respective regions. Two analysis methods were utilized to estimate the X-ray polarization of 1ES 1101--232: (a) a model-independent method implemented in the \tt{PCUBE} algorithm within the \tt{xpbin} task of \tt{ixpeobssim} \citep{2015APh....68...45K}, and (b) a spectropolarimetric (hereafter SpecPol) analysis using \tt{Xspec} \citep[v.12.14.1;][]{1999ascl.soft10005A}, as introduced in \citet{2017ApJ...838...72S}. For the \tt{PCUBE} analysis, we excluded the influence of the sky background following the background-subtraction procedure provided in \citet{2022SoftX..1901194B}.

The X-ray polarization properties of 1ES 1101--232 were first measured using the \tt{PCUBE} algorithm applied to the IXPE data from all three DUs combined. This analysis yielded the following parameters: the minimum detectable polarization at the 99\% CL (MDP$_{99}$), the normalized Stokes parameters $q$ and $u$ (where $q=Q/I$ and $u=U/I$), the X-ray polarization degree $\Pi_{\rm X}$ and angle $\psi_{\rm X}$. The \tt{PCUBE} analysis was performed using the unweighted analysis method \citep[i.e., weights = False and irfname = "ixpe:obssim:20240701:v013";][]{2022AJ....163..170D}.

\begin{figure*}
    \centering
    \includegraphics[angle=0, scale=0.45]{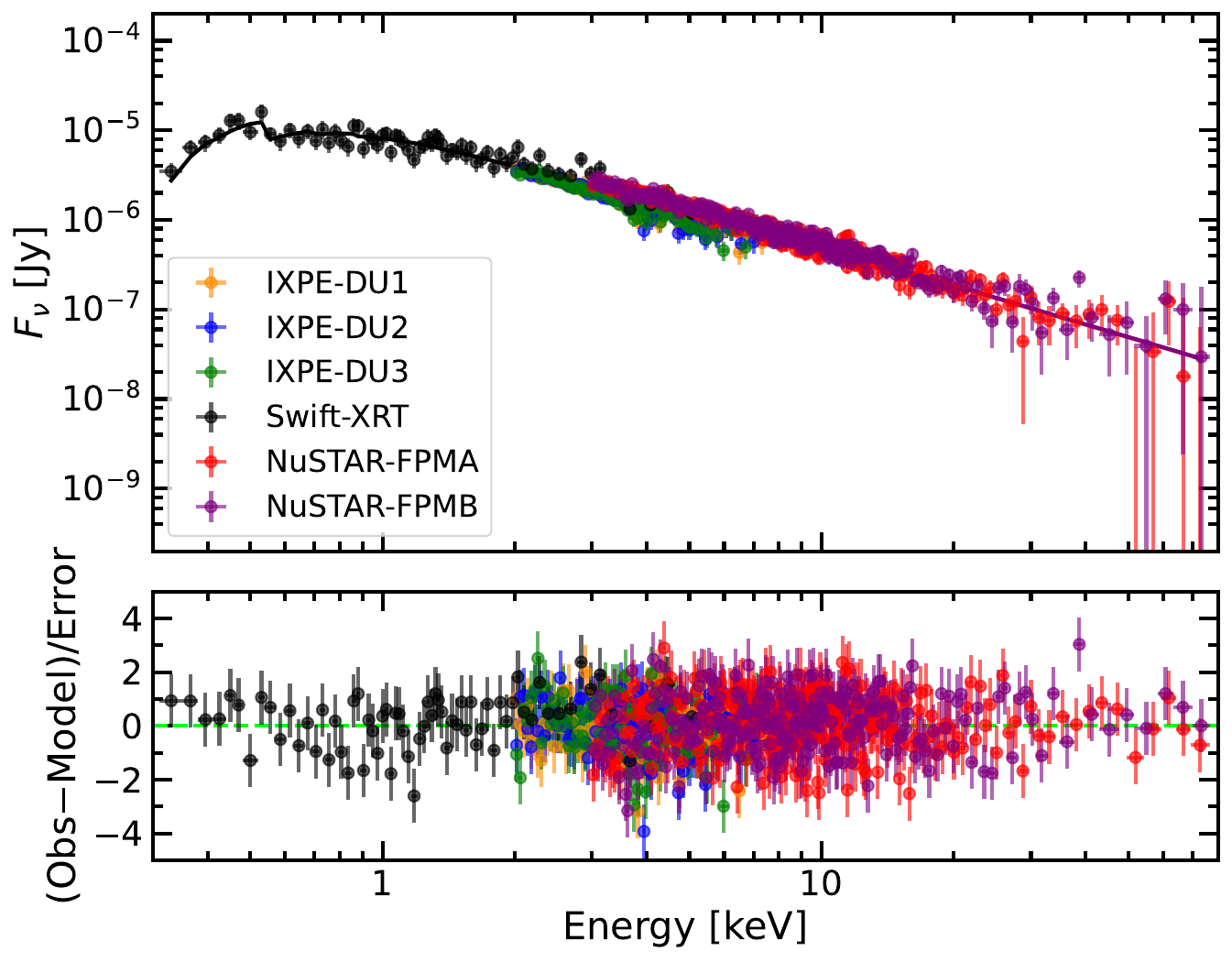}
    \caption{Broadband X-ray spectrum of 1ES 1101--232 in the 0.3--79 keV band, which was jointly fitted using the contemporaneous from IXPE (orange, blue and green points), Swift-XRT (black points) and NuSTAR (red and violet points).}
    \label{fig:spec}
\end{figure*}

As a cross-check, the X-ray polarization of 1ES 1101--232 was then estimated through SpecPol fits in \tt{Xspec}, which is part of the \tt{HEASoft} package \citep[v.6.34;][]{2014ascl.soft08004N}. The spectra for the Stokes parameters $I$, $Q$ and $U$ from all three DUs were generated using the \tt{PHA1}, \tt{PHA1Q}, and \tt{PHA1U} algorithms, respectively, within the \tt{xpbin} task. The $I$ spectra were regrouped to ensure a minimum of 20 counts per energy bin, ensuring the validity of Gaussian statistics in the spectral fits. For the $Q$ and $U$ spectra, a constant energy width of 0.2 keV was employed. We used a weighted analysis method \citep[i.e., weights = True and irfname = "ixpe:obssim:20240701\_alpha075:v013";][]{2022AJ....163..170D} with the \tt{alpha075} response matrix files (RMFs) for the SpecPol fits to improve the significance of the estimates. The spectra were fitted following a two-step strategy \citep{2024ApJ...962...92X,2024ApJ...963....5E,2024ApJ...970L..22H,2025ApJ...986..182H}. In the first step, the $I$ spectra in the 2--8 keV band were jointly fitted with the simultaneous Swift-XRT and NuSTAR spectra (details will be described below) using an absorbed power-law (PL) model: \tt{CONSTANT}$\times$\tt{TBABS}$\times$\tt{POWERLAW} in \tt{Xspec} \citep{2025ApJ...986..182H}. The PL function is
\begin{equation}
\frac{dN}{dE}=N_{0}\times\left(\frac{E}{E_0}\right)^{-\Gamma},
\end{equation}
where $N_{0}$ is the PL normalization, $E_{0}=1~{\rm keV}$ is the scale parameter of photon energy, and $\Gamma$ is the photon spectral index \citep{2004A&A...413..489M}. In this model, the \tt{CONSTANT} and \tt{TBABS} components account for the uncertainties in the absolute effective areas of different detectors and the Galactic photoelectric absorption, respectively. During this procedure, $N_{0}$, the column density $N_{\rm H}$, and $\Gamma$ were left free to vary. Such a joint spectral fit using data from different detectors provides tighter constraints on $N_{\rm H}$ and the spectral shape of the target \citep{2024ApJ...963....5E}. The broadband X-ray spectrum is shown in Figure \ref{fig:spec}. In the second step, the $I$, $Q$, and $U$ spectra from all three DUs were fitted simultaneously with an absorbed PL model that includes constant polarization: \tt{CONSTANT}$\times$\tt{TBABS}$\times$\tt{POLCONST}$\times$\tt{POWERLAW} in \tt{Xspec} \citep{2025ApJ...986..182H}. The \tt{POLCONST} component assumes a constant polarization across a specific energy range and has only two free parameters, i.e., $\Pi_{\rm X}$ and $\psi_{\rm X}$. When performing the SpecPol fit, all other spectral parameters were fixed at their best-fit values obtained from the first-step analysis, while $\Pi_{\rm X}$ and $\psi_{\rm X}$ were treated as free parameters.

\subsection{Swift-XRT}\label{subsec:xrt}

In this work, we analyzed data from five Swift-XRT observations of 1ES 1101--232, three of which were simultaneous with the IXPE pointing. All Swift-XRT observations were performed in the Photon Counting (PC) mode. The data were processed using the XRT Data Analysis Software (XRTDAS, v.3.7.0) within the \tt{HEASoft} package. Calibration was performed with the \tt{xrtpipeline} task, utilizing the calibration files from the Swift-XRT CALDB (v.20241028). Source and background spectra were extracted using the \tt{xselect} task. To eliminate pile-up effects, we extracted source photons from an annulus centered on the brightest pixel of the image, with inner and outer radii of $11\arcsec$--$13\arcsec$ and $47\arcsec$, respectively. Background photons were extracted from a larger concentric annulus with inner and outer radii of $71\arcsec$ and $142\arcsec$, respectively. Ancillary response files (ARFs) were generated using the \tt{xrtmkarf} task, which incorporates cumulative exposure maps to account for point-spread function (PSF) losses and CCD defects.

We merged the event files from the three observations simultaneous with the IXPE pointing and generated a combined spectrum using the \tt{xselect} task. A combined exposure map was also created from the cumulative exposure maps of these three observations using the \tt{ximage} task, which was then used to generate the corresponding ARF for the combined spectrum. This combined Swift-XRT spectrum was jointly fitted with the IXPE $I$ spectra in the first step of the SpecPol fit to enhance the signal-to-noise ratio (S/N) and to precisely constrain the spectral parameters.

\subsection{NuSTAR}\label{subsec:nustar}
NuSTAR, equipped with two co-aligned multilayer-coated telescopes (FPMA and FPMB), observed 1ES 1101--232 from 12:01 UTC on 2024 November 29 to 07:01 UTC on 2024 December 1, achieving a net exposure of $\sim35$ ks (OBSID: 61001017002; PI: L. Ballo). This observation was simultaneous with the IXPE pointing. The raw data were processed using the \tt{nupipeline} task, which is implemented in the NuSTAR Data Analysis Software (NuSTARDAS, v.2.1.4) within the \tt{HEASoft} package. Calibrated and cleaned event files were generated with the \tt{nupipeline} task, utilizing calibration files from the NuSTAR CALDB (v.20250602). The source region was selected as a circle with a radius of $90\arcsec$ centered on the centroid of the X-ray emission. A background region of the same size was chosen from a nearby, source-free area on the image. Source and background spectra, along with the corresponding RMFs and ARFs, were generated using the \tt{nuproducts} task. Subsequently, the NuSTAR spectra were jointly fitted with the IXPE $I$ spectra in the first step of the SpecPol fit for the same reason as described in Section \ref{subsec:xrt}.

\section{Results of the X-ray Polarization Observations}\label{sec:pol}
\subsection{Time- and Energy-averaged Analysis}\label{subsec:aver}

The X-ray polarization of 1ES 1101--232 was first measured in the 2--8 keV energy band over the entire IXPE pointing epoch using both the \tt{PCUBE} algorithm and the SpecPol fit in \tt{Xspec}. From the \tt{PCUBE} analysis, a polarization degree of $\Pi_{\rm X}=17.3\%\pm3.7\%$ and a polarization angle of $\psi_{\rm X}=14\degr.5\pm6\degr.1$ are obtained at the 4.7$\sigma$ CL, with a MDP$_{99}$ of $\sim11.2\%$. From the SpecPol fit, the X-ray polarization parameters are $\Pi_{\rm X}=16.0\%\pm2.6\%$ and $\psi_{\rm X}=10\degr.7\pm4\degr.7$ at the 6.2$\sigma$ CL. However, the effective area of IXPE decreases rapidly above 6 keV (see Figure 5.1 in the IXPE Observatory User Guide\footnote{\url{https://heasarc.gsfc.nasa.gov/docs/ixpe/analysis/IXPE-SOC-DOC-011A_UG-Observatory.pdf}}). Since photon events in the 6--8 keV band do not significantly enhance the precision of the polarization measurements, we re-performed the polarization analysis within the 2--6 keV energy range instead of the full IXPE band. This strategy was also applied in \citet{2025MNRAS.540.3242D}. In this case, the \tt{PCUBE} analysis yields $\Pi_{\rm X}=19.3\%\pm3.1\%$ and $\psi_{\rm X}=12\degr.4\pm4\degr.6$ at the 6.2$\sigma$ CL in the 2--6 keV band, with a MDP$_{99}$ of $\sim9.4\%$, while $\Pi_{\rm X}=17.9\%\pm2.7\%$ and $\psi_{\rm X}=10\degr.0\pm4\degr.4$ at the 6.6$\sigma$ CL are derived through the SpecPol fit. Evidently, the statistical significance of the polarization measurements in the 2--8 keV band is lower than that in the 2--6 keV band. Therefore, the X-ray polarization properties of 1ES 1101--232 in the 2--6 keV band are selected for the subsequent discussion.

\begin{figure*}
    \centering
    \begin{minipage}{0.45\textwidth}
        \centering
        \includegraphics[angle=0, height=\textwidth]{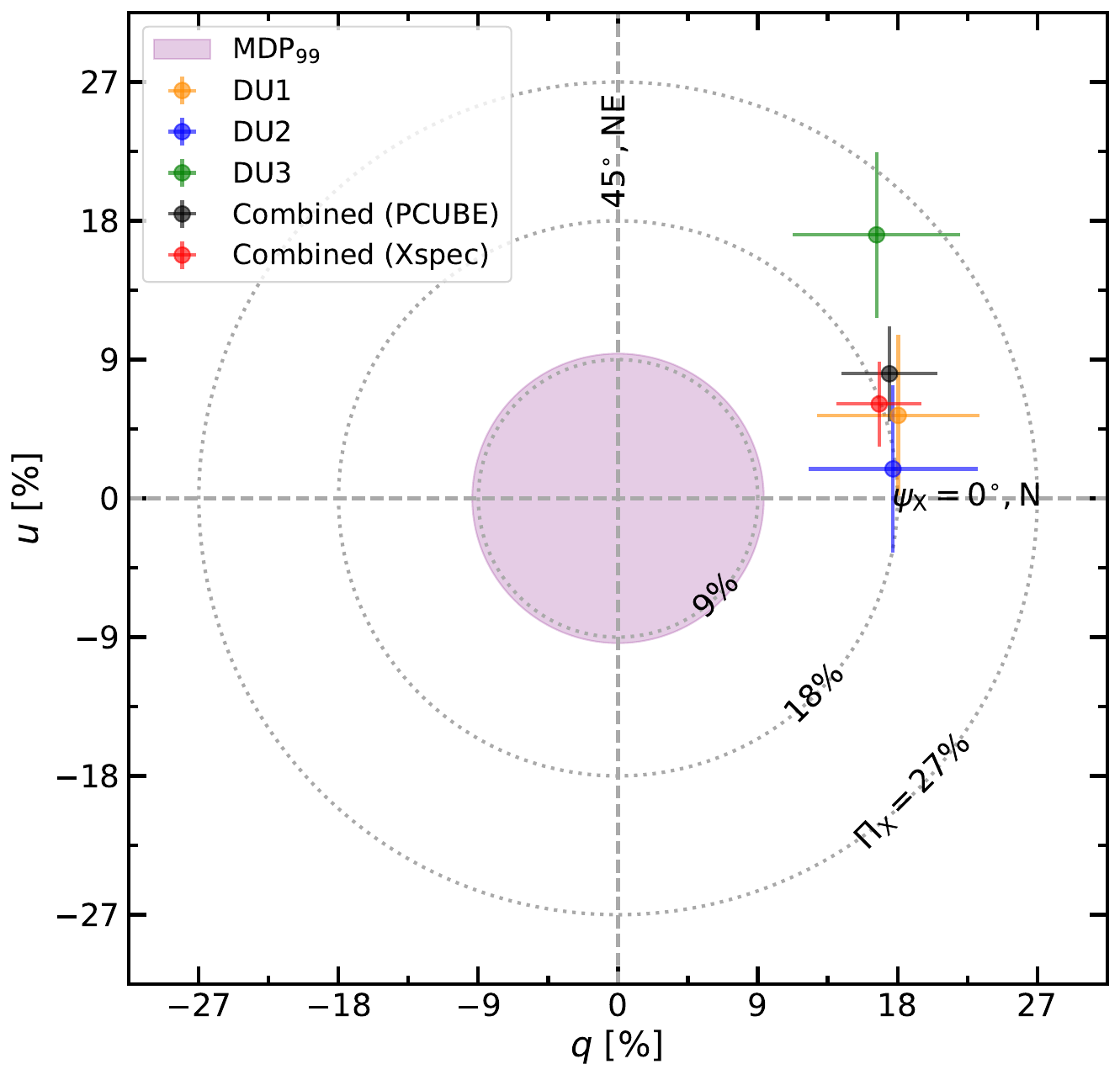} \\
            \makebox{(a)}
    \end{minipage}
    \begin{minipage}{0.45\textwidth}
        \centering
        \includegraphics[angle=0, height=\textwidth]{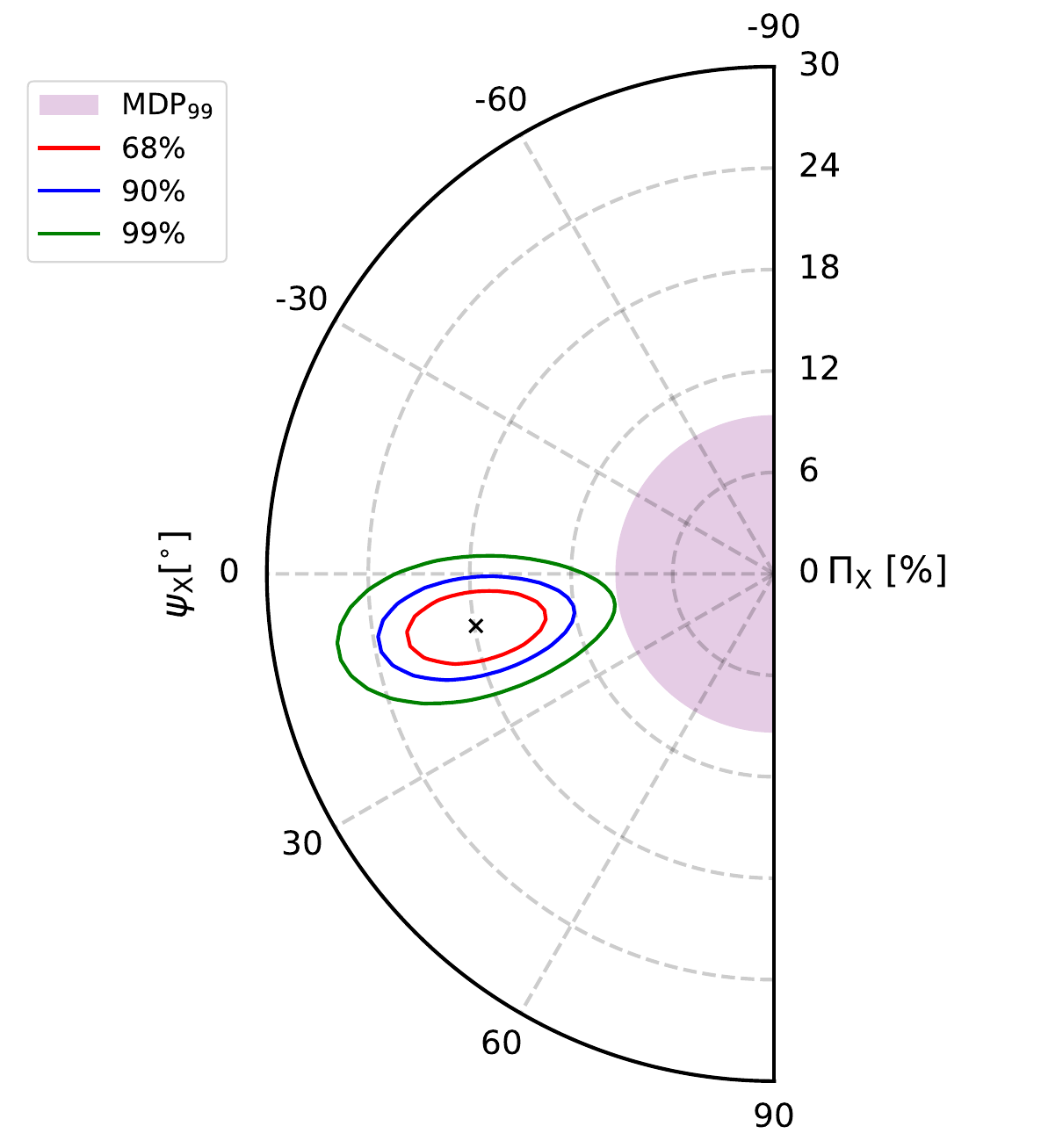} \\
            \makebox{(b)}
    \end{minipage}
    \caption{Normalized Stokes parameters $q$ and $u$ (sub-figure [a]) and 2-D polarization contours between $\Pi_{\rm X}$ and $\psi_{\rm X}$ (sub-figure [b]) of 1ES 1101--232 in the 2--6 keV band. In sub-figure (a), the orange, blue, green, and black points represent the results of DU1, DU2, DU3, and the combination of the three DUs, respectively, measured through the \tt{PCUBE} algorithm within \tt{ixpeobssim}. The red point represents the result of the combination of the three DUs, estimated via SpecPol fit within \tt{Xspec}. In sub-figure (b), the black cross marks the best-fit values of $\Pi_{\rm X}$ and $\psi_{\rm X}$ and the red, blue and green contours refer to the results at the 68\%, 90\% and 99\% CLs, respectively, derived through the SpecPol fit. The violet shaded areas in both sub-figures represent the MDP$_{99}$ value of the IXPE observation for 1ES 1101--232.}
    \label{fig:qu&cont}
\end{figure*}

\begin{figure*}
    \centering
    \includegraphics[angle=0, scale=0.4]{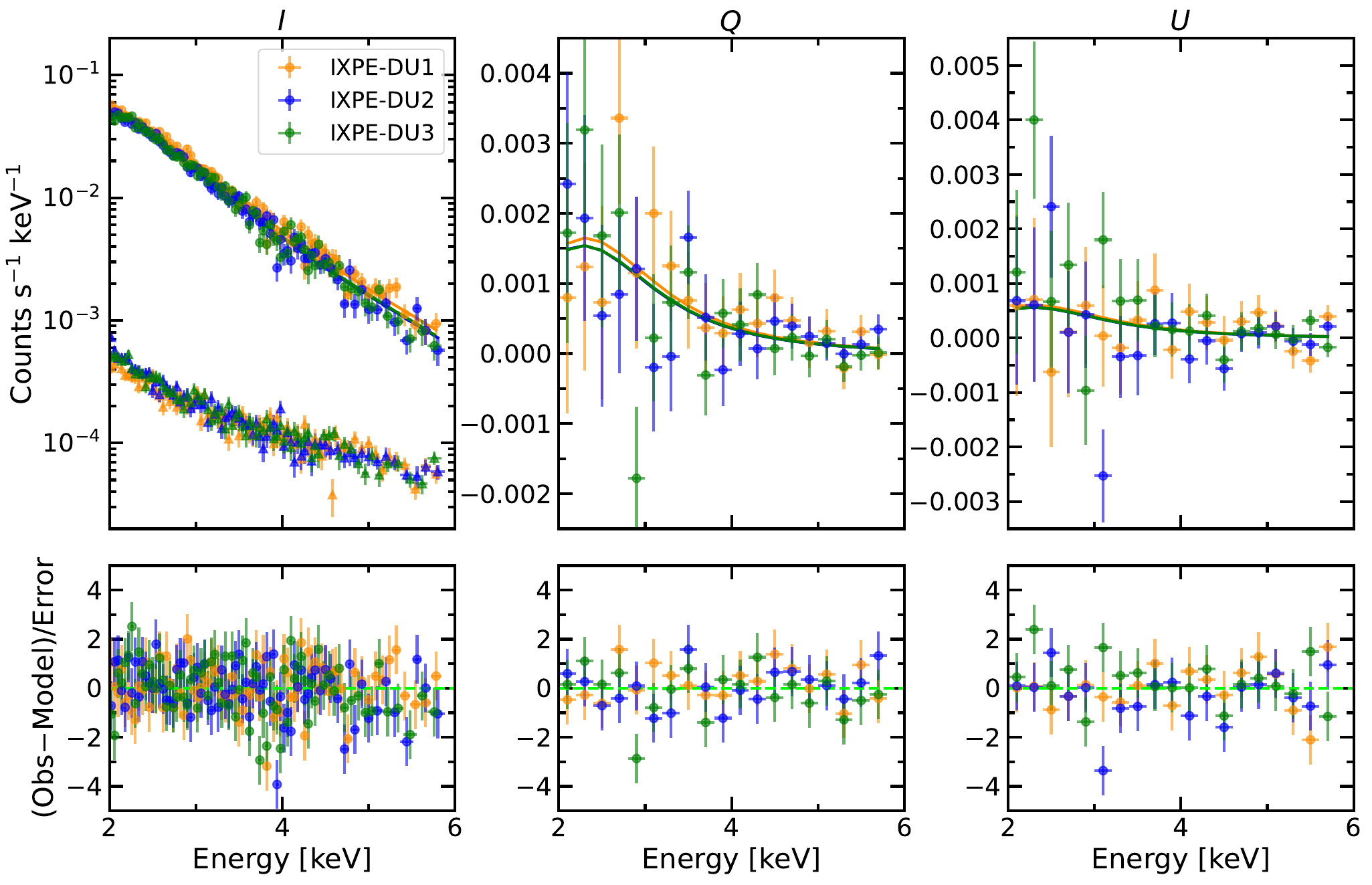}
    \caption{SpecPol fit result of 1ES 1101--232 in the 2--6 keV band. The panels from left to right represent the fits to the IXPE Stokes parameter $I$, $Q$ and $U$ with the associated residuals. In the $I$ spectra panel, the triangles indicate the background spectra for different DUs.}
    \label{fig:specpol}
\end{figure*}

The normalized Stokes parameters derived within the 2--6 keV band for 1ES 1101--232 are shown in Figure \ref{fig:qu&cont}. The 2-D polarization contours of $\Pi_{\rm X}$ and $\psi_{\rm X}$ at the 68\%, 90\%, and 99\% CLs, which were estimated using the \tt{steppar} task in \tt{Xspec}, are also present in Figure \ref{fig:qu&cont}. The X-ray polarization parameters of 1ES 1101--232 derived from the \tt{PCUBE} analysis and the SpecPol fit are statistically consistent within their 1$\sigma$ uncertainties. However, the best-fit values of $\Pi_{\rm X}$ and $\psi_{\rm X}$ from these two methods are not identical. This discrepancy can be explained by the fundamental difference between the approaches: the \tt{PCUBE} algorithm provides model-independent polarization parameters, while the SpecPol fit incorporates the constraints from the best-fit spectral model \citep{2024A&A...681A..12K}. Given that the weighted analysis method used in the SpecPol fits may enhance the sensitivity of the polarization measurement (see \citealt{2022AJ....163..170D} for details), we will primarily use the SpecPol results for the subsequent description and discussion of the X-ray polarization properties of 1ES 1101--232. The best-fit parameters of the SpecPol fit are summarized in Table \ref{tab:specpol}, and the SpecPol fit results in the 2--6 keV band for the $I$, $Q$, and $U$ spectra are presented in Figure \ref{fig:specpol}.

Additionally, in the first step of the SpecPol fit, the joint spectral fit yielded a column density of $N_{\rm H}=1.34^{+0.11}_{-0.10}\times10^{21}~{\rm cm^{-2}}$ and a photon index of $\Gamma=2.45\pm0.01$ for the broadband X-ray spectrum of 1ES 1101--232. These values indicate that the source was highly absorbed and exhibited a soft X-ray spectrum. The unabsorbed flux in the 2--8 keV band ($F_{\rm 2-8}$) during the IXPE observation was calculated to be $(1.96\pm0.05)\times10^{-11}~{\rm erg~cm^{-2}~s^{-1}}$ using the \tt{cflux} task within \tt{Xspec}. The best-fit parameters from the joint spectral fit are also summarized in Table \ref{tab:specpol}.

\subsection{Temporal Analysis}\label{subsec:tbin}

\begin{figure*}
    \centering
    \includegraphics[angle=0, scale=0.45]{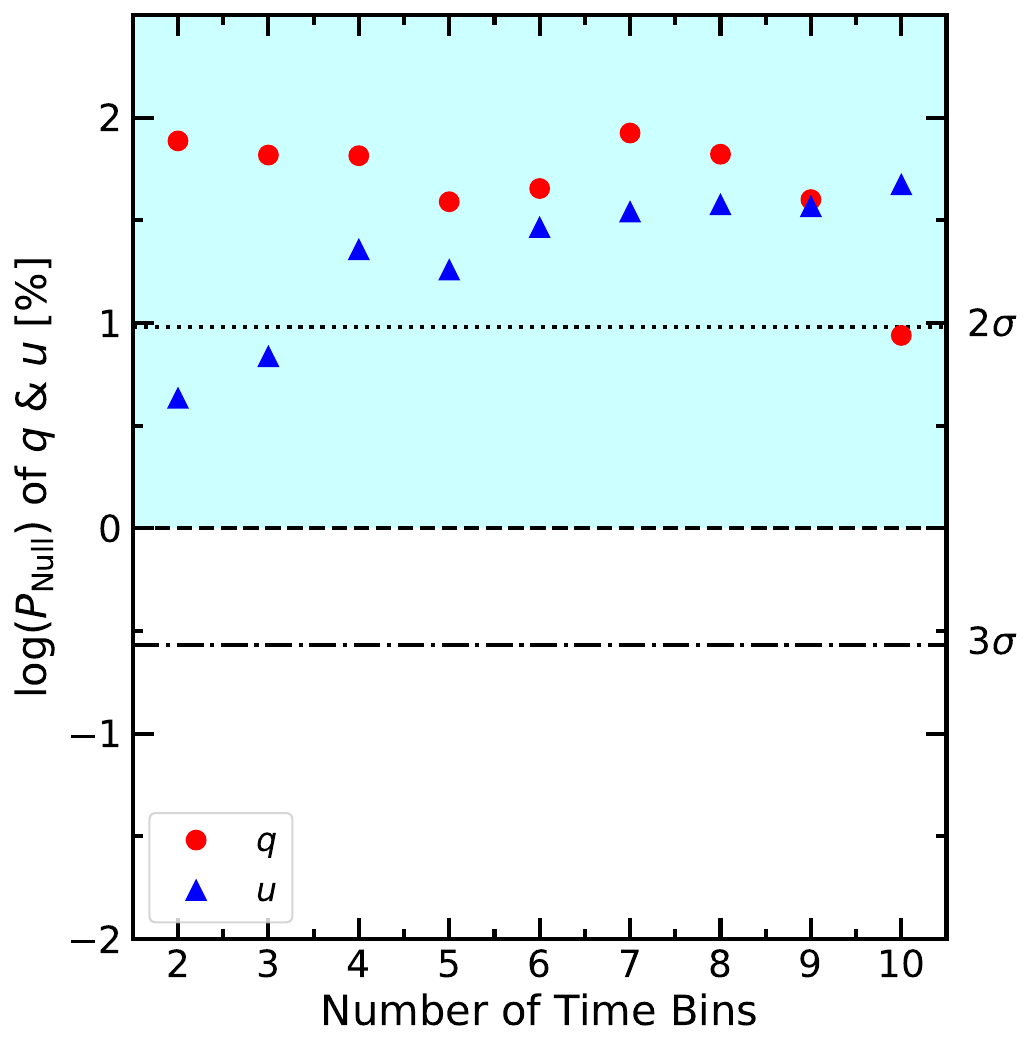}
    \caption{$P_{\rm Null}$ of $\chi^{2}$ test on the variation of the normalized Stokes parameters $q$ (red points) and $u$ (blue points) in logarithmic scale as a function of number of time bins, assuming that the X-ray polarization of 1ES 1101--232 remained constant druing the IXPE pointing epoch. The cyan shaded area represent the region where $P_{\rm Null}>1\%$. The dotted and dash-dot lines represent the statistical significance where the null hypothesis can be rejected at the 2$\sigma$ (90.45\%) and 3$\sigma$ (99.73\%) CLs.}
    \label{fig:pnull}
\end{figure*}

To investigate potential temporal variations in the X-ray polarization of 1ES 1101--232, we performed a time-resolved analysis of the IXPE dataset following the procedures described in \citet{2024A&A...681A..12K}, \citet{2024ApJ...970L..22H}, and \citet{2025ApJ...986..182H}. We assessed temporal variability by calculating the null hypothesis probability ($P_{\rm Null}$) using $\chi^{2}$ test, under the assumption that both $\Pi_{\rm X}$ and $\psi_{\rm X}$ remained constant throughout the entire observation. For this test, we used the normalized Stokes parameters $q$ and $u$ derived from the \tt{PCUBE} analysis. We divided the complete dataset into two to ten time bins and estimated $q$ and $u$ for each bin. The $\chi^{2}$ statistic and corresponding $P_{\rm Null}$ were then calculated for each case. Figure \ref{fig:pnull} shows the results of this time-resolved analysis for 1ES 1101--232. Typically, $P_{\rm Null}<1\%$ suggests possible temporal variability in the X-ray polarization, while $P_{\rm Null}>1\%$ indicates statistical stability over the total exposure \citep{2024A&A...681A..12K,2024ApJ...970L..22H,2025ApJ...986..182H}. As shown in Figure \ref{fig:pnull}, the X-ray polarization of 1ES 1101--232 exhibits a hint of temporal variation at a CL exceeding 2$\sigma$ when the observation is divided into two, three, and ten bins.

\begin{figure*}
    \centering
    \begin{minipage}{0.45\textwidth}
        \centering
        \includegraphics[angle=0, width=\textwidth]{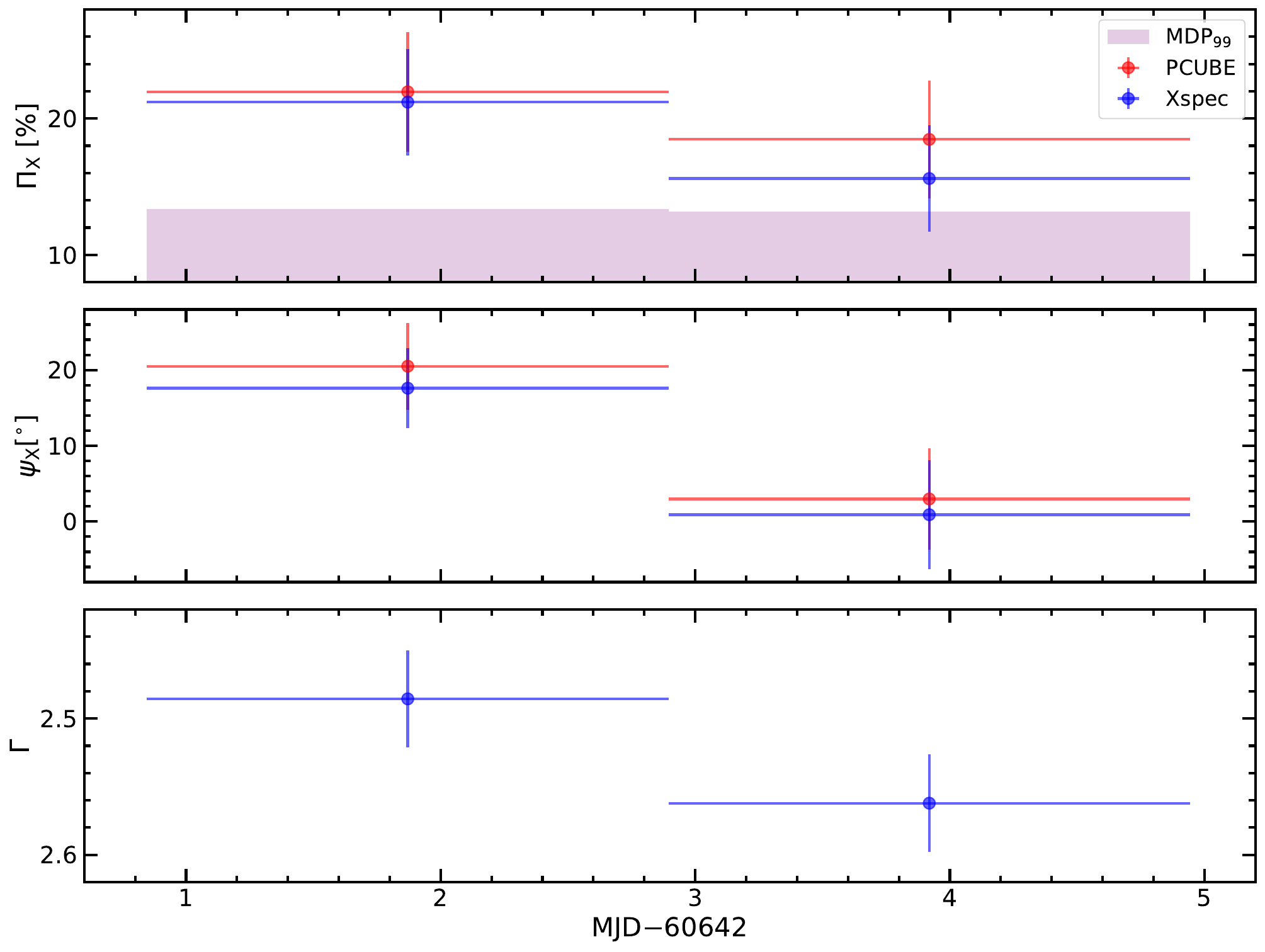} \\
            \makebox{(a)}
    \end{minipage} \\
    \begin{minipage}{0.45\textwidth}
        \centering
        \includegraphics[angle=0, width=\textwidth]{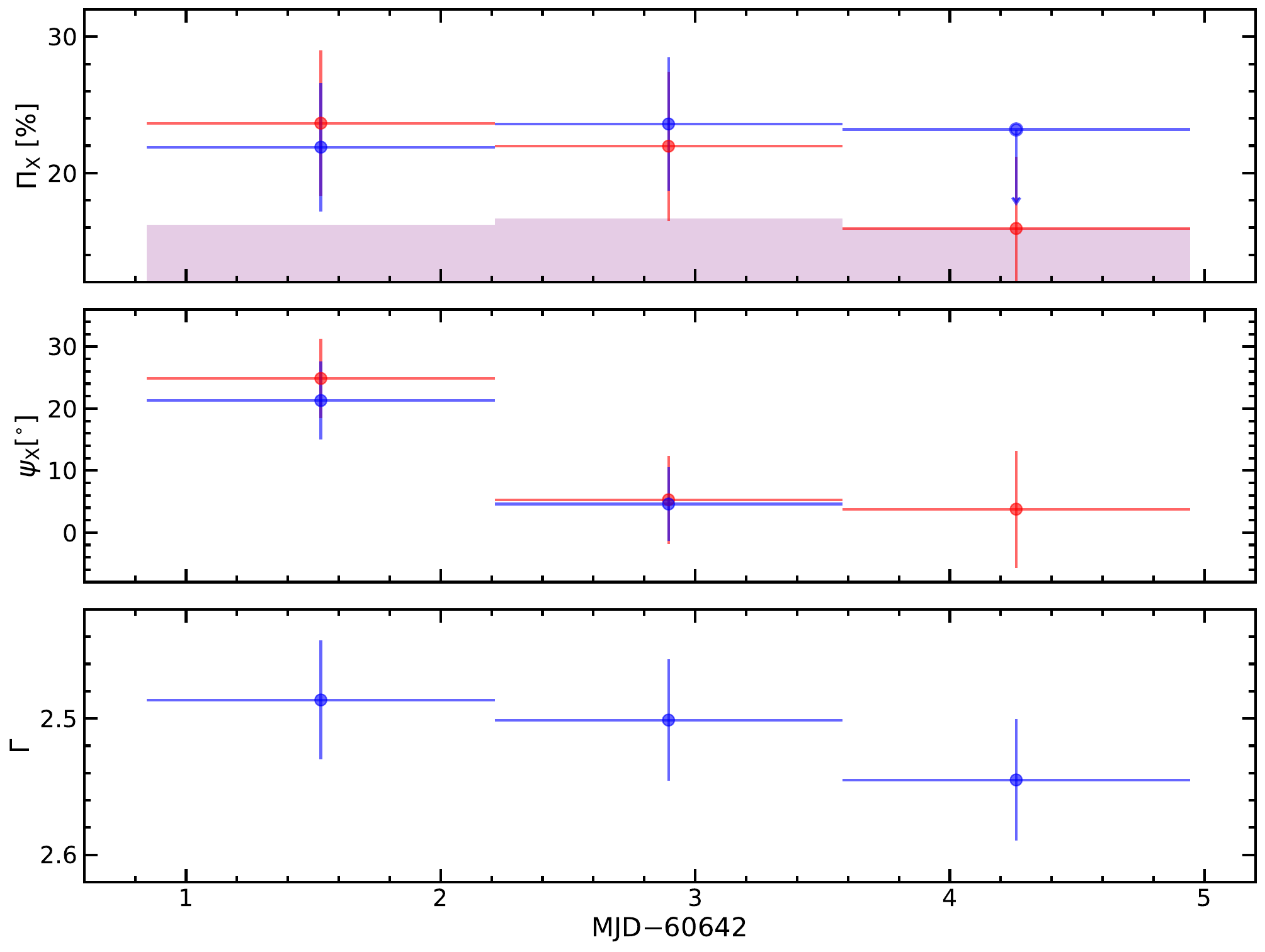} \\
            \makebox{(b)}
    \end{minipage} \\
    \begin{minipage}{0.45\textwidth}
        \centering
        \includegraphics[angle=0, width=\textwidth]{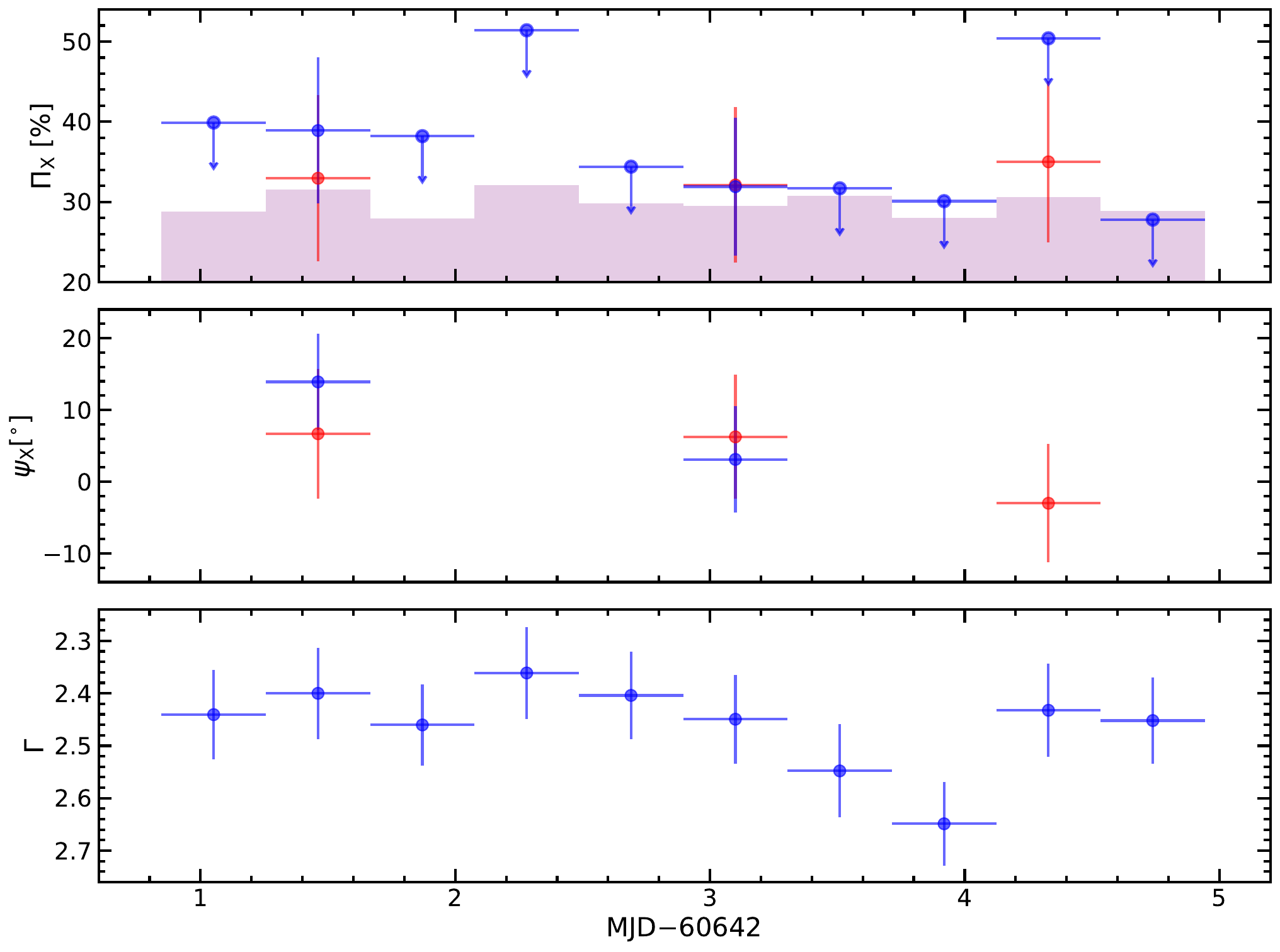} \\
            \makebox{(c)}
    \end{minipage}
    \caption{Polarization light curves of 1ES 1101--232 in the 2--6 keV band. In each sub-figure, the top, middle and bottom panels represent light curves of $\Pi_{\rm X}$, $\psi_{\rm X}$ and $\Gamma$, respectively. The red and blue points signify the results derived via the \tt{PCUBE} analysis and SpecPol fits, respectively. In the $\Pi_{\rm X}$ panels, the violet shaded histograms donate the values of MDP$_{99}$ for different time bins, and the blue points with downward arrows mark the 1-D upper limits of $\Pi_{\rm X}$ at the 99\% CL, estimated using the \tt{error} task within \tt{Xspec} when the best-fit values are lower than the associated values of MDP$_{99}$. Sub-figures (a), (b) and (c) correspond to the cases that the IXPE observation is divided into two, three and ten time bins, respectively.}
    \label{fig:tbin}
\end{figure*}

To further explore the temporal variability of X-ray polarization, we constructed polarization light curves for 1ES 1101--232 by dividing the IXPE observation into two, three, and ten time bins, based on time-resolved analysis results for each bin $T_{\rm n,i}$ (where $n$ denotes the total number of intervals and $i$ indicates the sequential order of the interval), as shown in Figure \ref{fig:tbin}. For the cases with two and three time bins, there is a slight tendency for variation in both $\Pi_{\rm X}$ and $\psi_{\rm X}$, accompanied by a softening of the spectrum. In the case of ten time bins, the $\Pi_{\rm X}$ measurements primarily consist of upper limits, probably attributed to the substantially fewer photon counts in each time bin compared to the two- and three-bin cases. Nonetheless, strong polarization was detected in time bin $T_{10,2}$ through the SpecPol fit, with $\Pi_{\rm X}=38.9\%\pm9.1\%$ and $\psi_{\rm X}=13\degr.9\pm6\degr.7$ at the 4.3$\sigma$ CL. Additionally, the temporal evolution of $\psi_{\rm X}$ in the ten-bin case shows a trend similar to that observed in the two- and three-bin cases, exhibiting a slight decrease throughout the IXPE observation.

\begin{figure*}
    \centering
    \begin{minipage}{0.3\textwidth}
        \centering
        \includegraphics[angle=0, width=\textwidth]{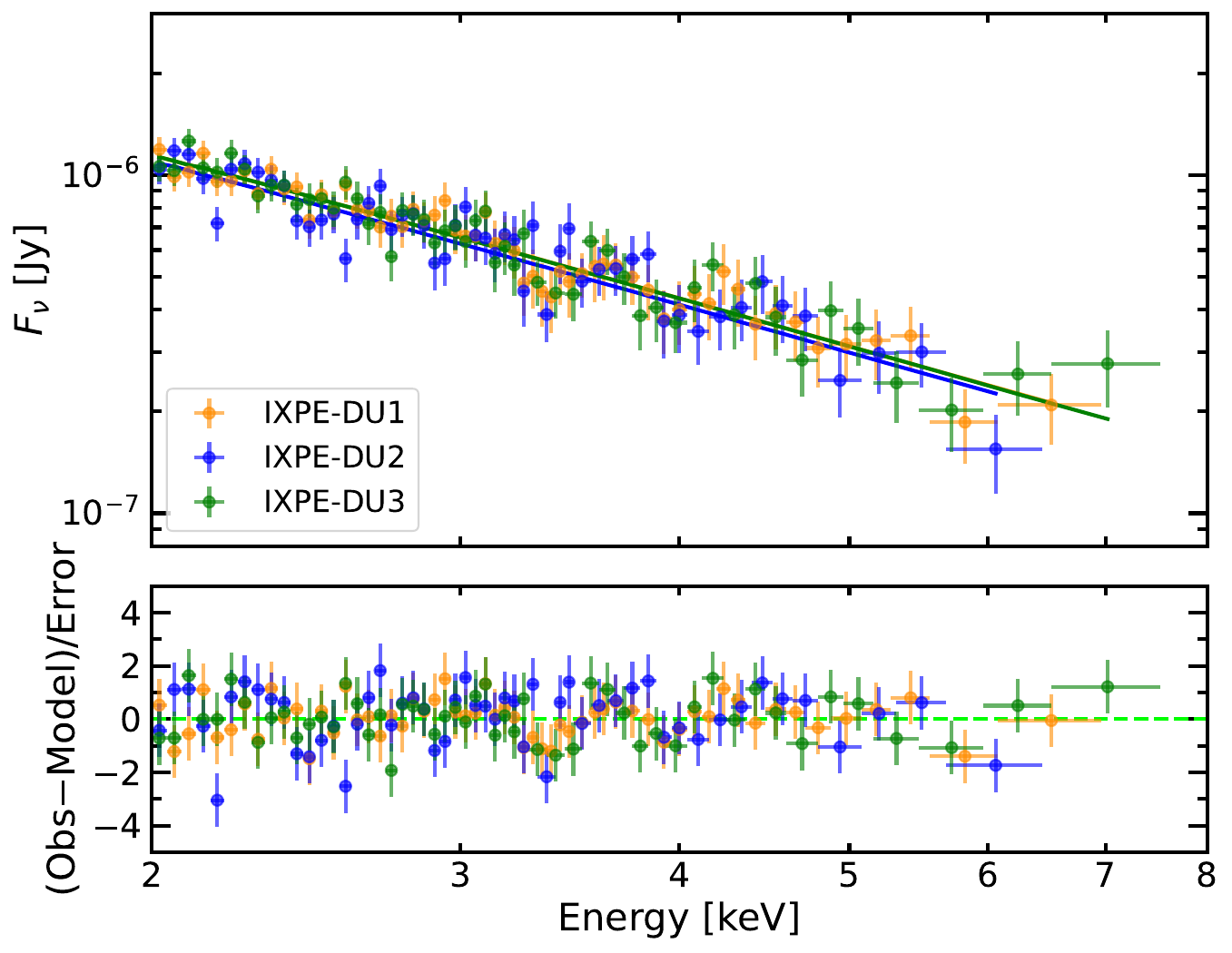} \\
            \makebox{(a)}
    \end{minipage}
    \begin{minipage}{0.3\textwidth}
        \centering
        \includegraphics[angle=0, width=\textwidth]{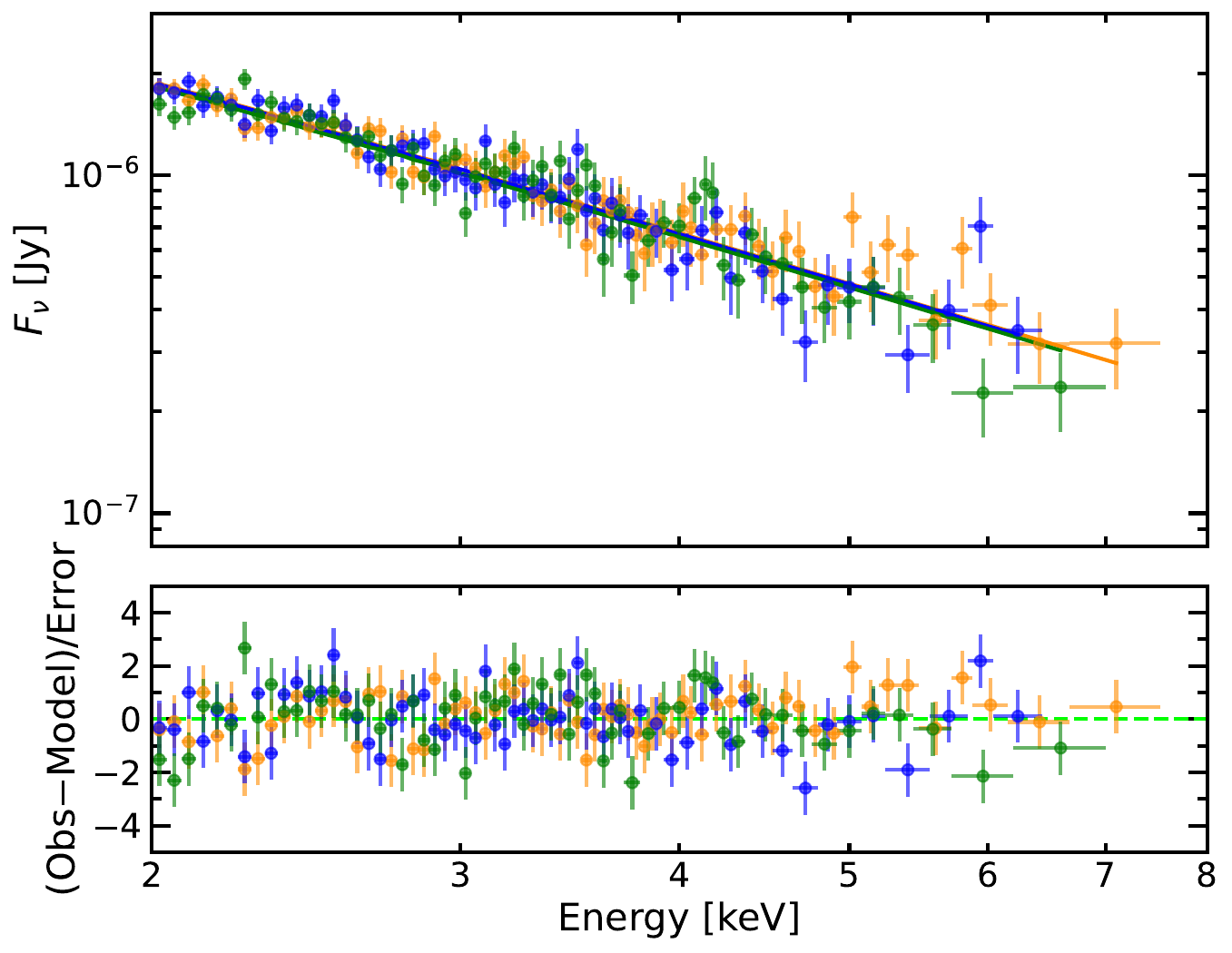} \\
            \makebox{(b)}
    \end{minipage}
    \begin{minipage}{0.3\textwidth}
        \centering
        \includegraphics[angle=0, width=\textwidth]{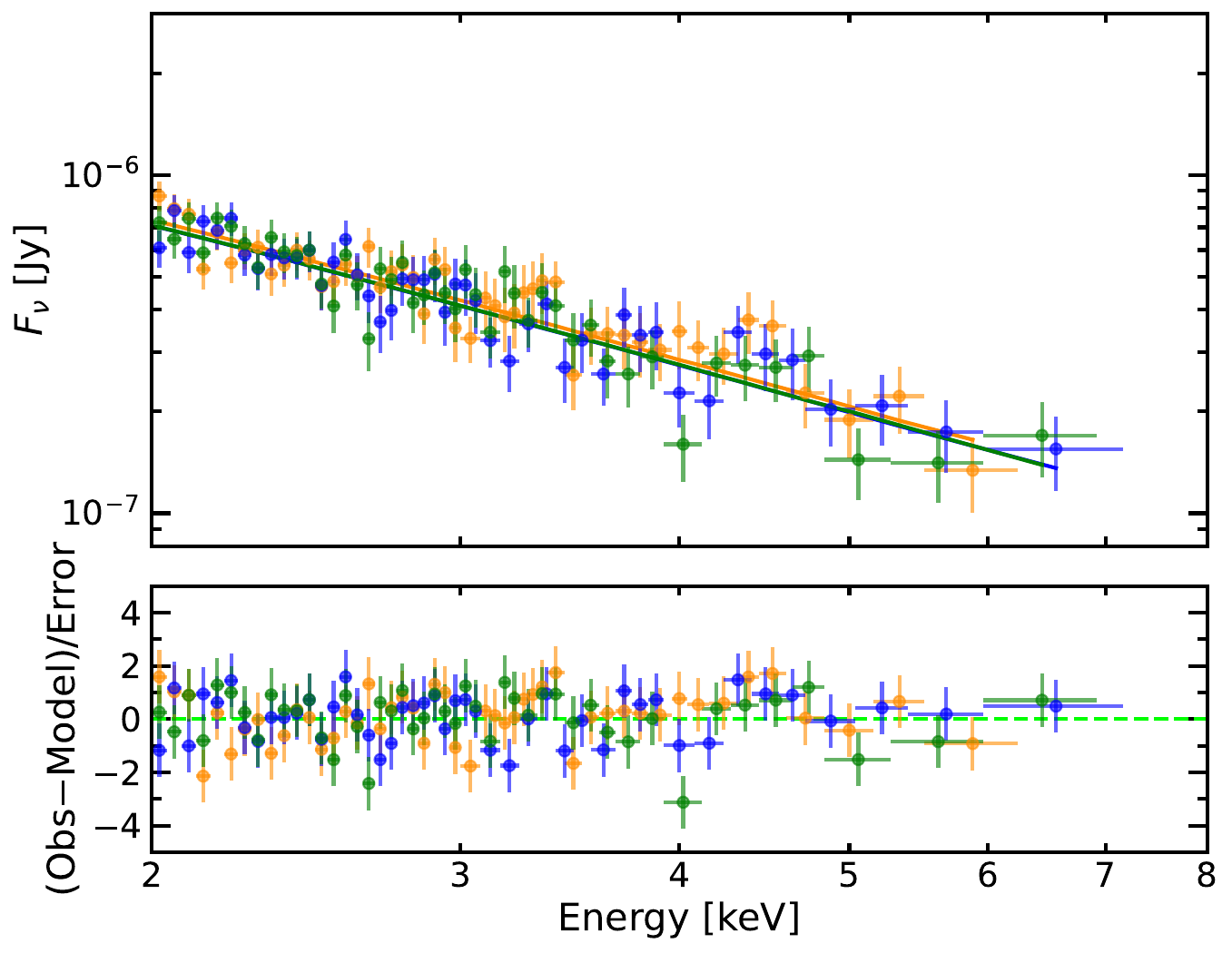} \\
            \makebox{(c)}
    \end{minipage}
    \caption{The IXPE $I$ spectra for the three custom time intervals ($T_{\rm c,1}$ (a), $T_{\rm c,2}$ (b), and $T_{\rm c,3}$ (c)) in the 2--8 keV band.}
    \label{fig:spec-tbin}
\end{figure*}

Interestingly, the aforementioned ten-bin segmentation reveals that the spectra of 1ES 1101--232 exhibit three distinct phases: a stable phase (from $T_{10,1}$ to $T_{10,3}$), a softening phase (from $T_{10,4}$ to $T_{10,8}$), and a hardening phase returning to levels comparable to the initial state (from $T_{10,9}$ to $T_{10,10}$), as displayed in Figure \ref{fig:tbin}. Based on these spectral variations, we divided the total IXPE exposure into three custom time bins ($T_{\rm c,1}$, $T_{\rm c,2}$, and $T_{\rm c,3}$) corresponding to these three phases, enabling a more detailed investigation of the temporal evolution of X-ray polarization. Firstly, the IXPE $I$ spectra during the three custom time intervals were produced, as demonstrated in Figure \ref{fig:spec-tbin}, indicating the obvious spectral variation. The polarization analysis results for the three custom time bins are presented in Figure \ref{fig:custom}. Under this refined temporal segmentation, the X-ray polarization of 1ES 1101--232 shows a plausible variation trend in the 1-D light curve; that is, $\Pi_{\rm X}$ first decreases from $T_{\rm c,1}$ to $T_{\rm c,2}$, followed by an increase from $T_{\rm c,2}$ to $T_{\rm c,3}$, while $\psi_{\rm X}$ slightly shifts from East to North in the sky projection. However, the 2-D contours between $\Pi_{\rm X}$ and $\psi_{\rm X}$ at the 99\% CL for these three periods are nearly overlapping, which implies that the X-ray polarization of 1ES 1101--232 is statistically stable during the IXPE pointing epoch, as shown in Figure \ref{fig:custom}.

\subsection{Energy-dependent Analysis}\label{subsec:ebin}

We also tested for energy dependence in the X-ray polarization of 1ES 1101--232 by applying the same null-hypothesis test described in Section \ref{subsec:tbin}. The IXPE energy range (2--6 keV) was divided into progressively narrower bins: i.e., two bins (2--4 keV and 4--6 keV), three bins (2--3 keV, 3--4 keV, and 4--6 keV), and four bins (2--3 keV, 3--4 keV, 4--5 keV, and 5--6 keV). In all cases, the null-hypothesis tests are consistent with a constant polarization model, indicating that no significant energy dependence in the X-ray polarization of 1ES 1101--232 was found. The detailed results of the energy-resolved analysis are shown in Figure \ref{fig:ebin}.

\section{Spectral Variation}\label{sec:spec}

As demonstrated in Figures \ref{fig:tbin}, \ref{fig:spec-tbin} and \ref{fig:custom}, the X-ray emission of 1ES 1101--232 exhibits significant spectral evolution during the IXPE pointing epoch. To further explore the X-ray flux variability and spectral variation 1ES 1101--232, we reanalyzed five Swift-XRT observations of the source that were simultaneous or quasi-simultaneous with the IXPE observation. Notably, the two quasi-simultaneous Swift-XRT observations were strategically timed, one immediately preceding and the other immediately following the IXPE observation, thereby enabling a more precise characterization of the source's spectral evolution in X-rays. The data reduction procedures are described in Section \ref{subsec:xrt}, and the corresponding best-fit parameters are summarized in Table \ref{tab:xrt}.

As shown in Figure \ref{fig:lc}, the unabsorbed flux in the 0.3--10 keV band ($F_{0.3-10}$) remains statistically constant, with the null hypothesis probability $P_{\rm Null}>90\%$ derived from $\chi^{2}$ test. The unabsorbed fluxes in the 0.3--2 keV ($F_{0.3-2}$, soft component) and 2--10 keV ($F_{2-10}$, hard component) bands are also presented in Figure \ref{fig:lc}. In contrast to $F_{0.3-10}$, $F_{2-10}$ shows variability at the 4.8$\sigma$ CL, whereas $F_{0.3-2}$ displays marginal variability at the 2.3$\sigma$ CL. Additionally, the soft component dominates the X-ray spectrum of 1ES 1101--232 throughout the Swift-XRT monitoring period. We calculated the hardness ratio (HR), defined as $(H-S)/(H+S)$, for each Swift-XRT observation, where $H$ and $S$ represent $F_{2-10}$ and $F_{0.3-2}$, respectively. The HR light curve, presented in Figure \ref{fig:lc}, indicates that the X-ray spectrum of 1ES 1101--232 initially softens and subsequently hardens during the Swift-XRT monitoring campaign, a trend consistent with that observed in the IXPE temporal analysis.
 
We present spectral hardness-intensity diagrams (HIDs) that depict $\Gamma$ as a function of $F_{0.3-10}$, $F_{0.3-2}$, and $F_{2-10}$ in Figure \ref{fig:flux-gamma}. To estimate the correlation coefficients $r$ between $\Gamma$ and fluxes in different energy bands while accounting for parameter uncertainties, we employed the bootstrap method \citep{1979AnSta...7....1E}, following the approach used in \citet{2025ApJ...986..182H}. The bootstrap analysis yielded $r=0.16$, $r=0.78$, and $r=-0.83$ for the $F_{0.3-10}-\Gamma$, $F_{0.3-2}-\Gamma$, and $F_{2-10}-\Gamma$ relations, respectively. These results suggest that $F_{0.3-10}$ is not significantly correlated with $\Gamma$, whereas emission in the 0.3--2 keV band shows a {\it softer-when-brighter} trend, and emission in the 2--10 keV band exhibits a {\it harder-when-brighter} trend. When considering the axis orientation in Figure \ref{fig:flux-gamma}, clockwise hysteresis patterns are evident in the HIDs across all three energy bands.

\section{Discussion and Conclusions}\label{sec:disc&conc}

We present an analysis of the first IXPE observation of the EHSP 1ES 1101--232, conducted in conjunction with simultaneous Swift-XRT and NuSTAR observations. Significant X-ray polarization is detected in the 2--6 keV band at the 6.6$\sigma$ CL, with $\Pi_{\rm X}=17.9\%\pm2.7\%$ and $\psi_{\rm X}=10\degr.0\pm4\degr.4$. Within a narrower time interval, an even higher polarization is observed at the 4.3$\sigma$ CL, with $\Pi_{\rm X}=38.9\%\pm9.1\%$ and $\psi_{\rm X}=13\degr.9\pm6\degr.7$.

In addition to 1ES 0229+200 \citep{2023ApJ...959...61E} and H 1426+428 \citep{2025ApJ...986..182H}, 1ES 1101--232 represents the third EHSP observed by the IXPE. All three EHSPs exhibit significant X-ray polarization: H 1426+428 displays the highest polarization degree, with $\Pi_{\rm X}=20.6\%\pm2.9\%$, while both 1ES 0229+200 and 1ES 1101--232 show comparable values of $\Pi_{\rm X}\sim18\%$. During the IXPE observations, 1ES 1101--232 and H 1426+428 are characterized by PL X-ray spectra, while the spectrum of 1ES 0229+2000 requires a log-parabola model to describe and displays the hardest spectral index among the three sources, with $\Gamma\sim1.82$. Distinct spectral shapes may indicate different particle acceleration mechanisms. Curved spectra could be consistent with both stochastic acceleration and energy-dependent acceleration probability scenarios, while PL-shaped spectra might result from first-order relativistic shock acceleration \citep[e.g.,][]{2009A&A...501..879T}. These IXPE results suggest that the X-ray emission in these objects originates from a region with well-ordered magnetic field through synchrotron radiation. Furthermore, 1ES 0229+200 \citep{2023ApJ...959...61E} and H 1426+428 \citep{2025ApJ...986..182H} exhibit a chromatic behavior in polarization degree that is consistent with several typical HSPs, including Mrk 421 \citep{2022ApJ...938L...7D}, Mrk 501 \citep{2022Natur.611..677L}, PG 1553+113 \citep{2023ApJ...953L..28M}, and PKS 2155--304 \citep{2024A&A...689A.119K}. Moreover, Mrk 501 \citep{2022Natur.611..677L} and PKS 2155--304 \citep{2024A&A...689A.119K} exhibit a good alignment between their multiwavelength $\psi$ and their radio jet position angles. These behaviors can be naturally explained by an energy-stratified shock acceleration model \citep{1985ApJ...298..114M,2018MNRAS.480.2872T}, in which higher-energy particles lose energy more rapidly and radiate in close proximity to the acceleration site, where the magnetic field is ordered. Unfortunately, the absence of simultaneous polarimetric measurements at longer wavelengths for 1ES 1101--232 prevents us from further investigating its chromatic behavior of $\Pi$, the alignment of $\psi$ with its jet position angle \footnote{\citet{2012arXiv1205.2399T} conducted multiple observations of 1ES 1101--232 at 8 GHz from 2006 to 2009 by utilizing the National Radio Astronomy Observatory’s Very Long Baseline Array to investigate its parsec-scale structure. However, the jet position angle was not reported.}, and the alignment of $\psi$ among different wavelengths.

\begin{figure*}
    \centering
    \begin{minipage}{0.6\textwidth}
        \centering
        \includegraphics[angle=0, height=0.5\textwidth]{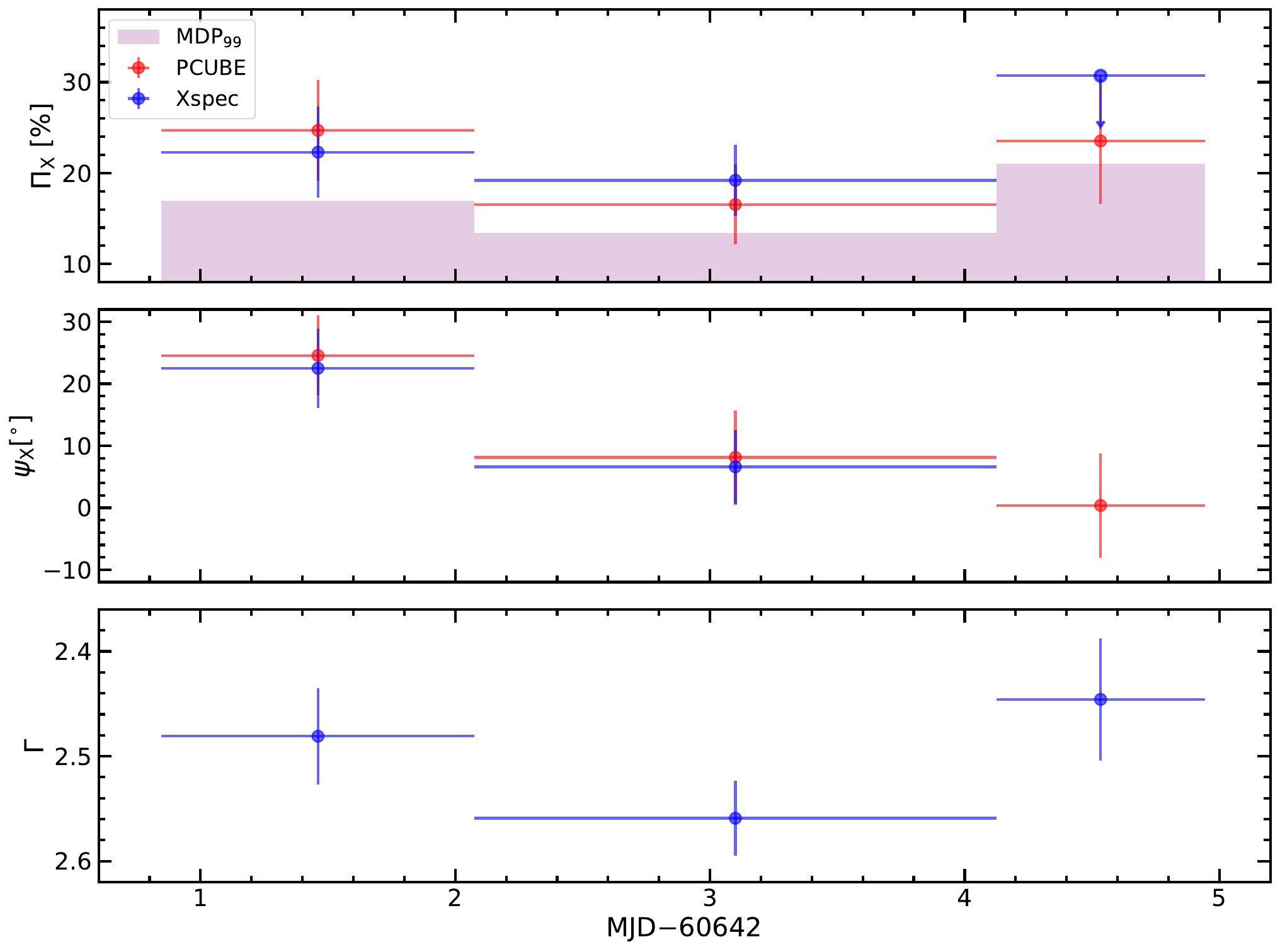} \\
            \makebox{(a)}
    \end{minipage}
    \begin{minipage}{0.3\textwidth}
        \centering
        \includegraphics[angle=0, height=\textwidth]{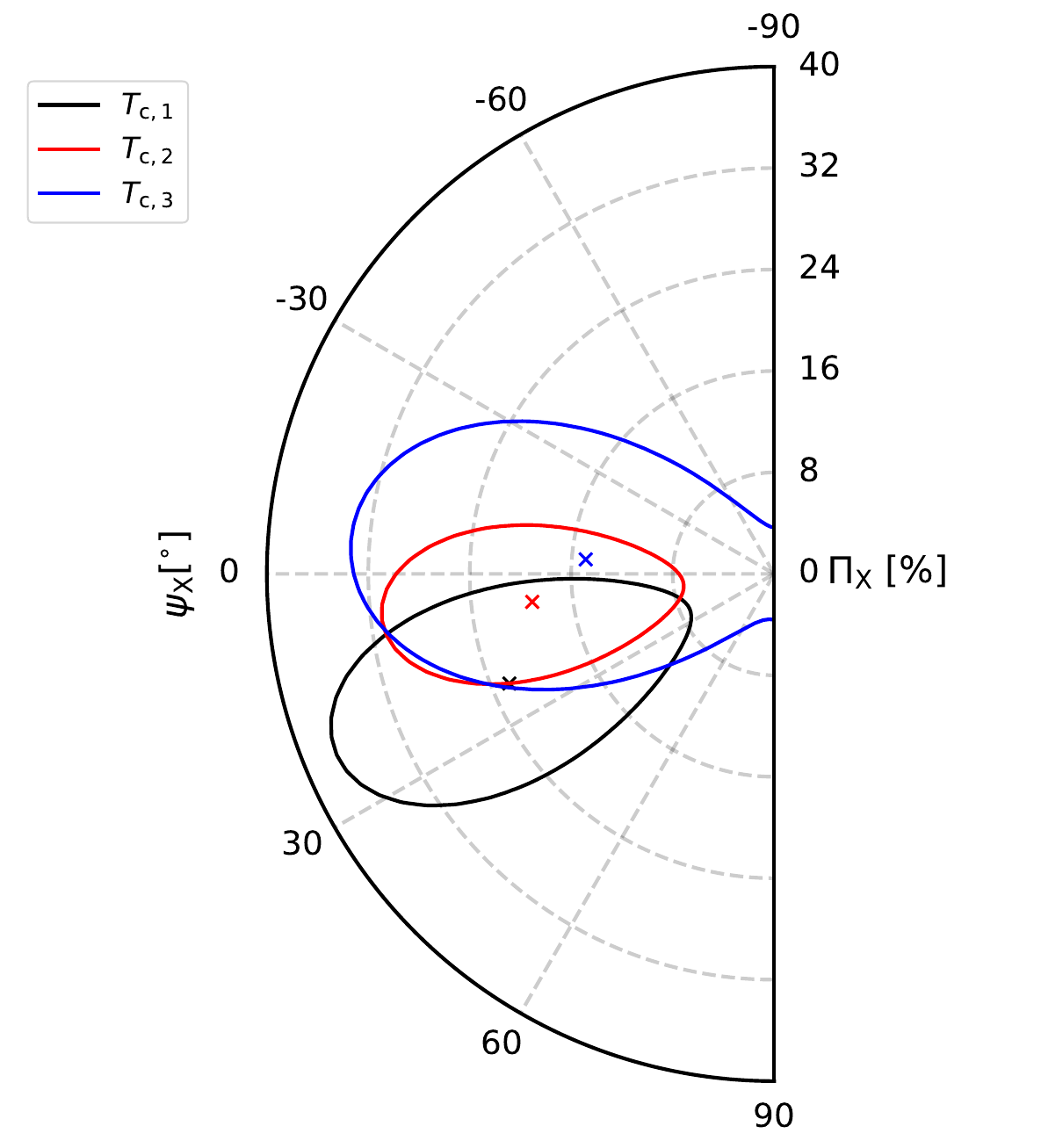} \\
            \makebox{(b)}
    \end{minipage}
    \caption{Polarization light curves (sub-figure [a]) and time-resolved polarization contours at the 99\% CL (sub-figure [b]) correspond to the case that the IXPE observation is divided into three custom time bins. Indications of the data points and shaded histogram in the left sub-figure are same with that utilized in Figure \ref{fig:tbin}.}
    \label{fig:custom}
\end{figure*}

\begin{figure*}
    \centering
    \begin{minipage}{0.45\textwidth}
        \centering
        \includegraphics[angle=0, width=\textwidth]{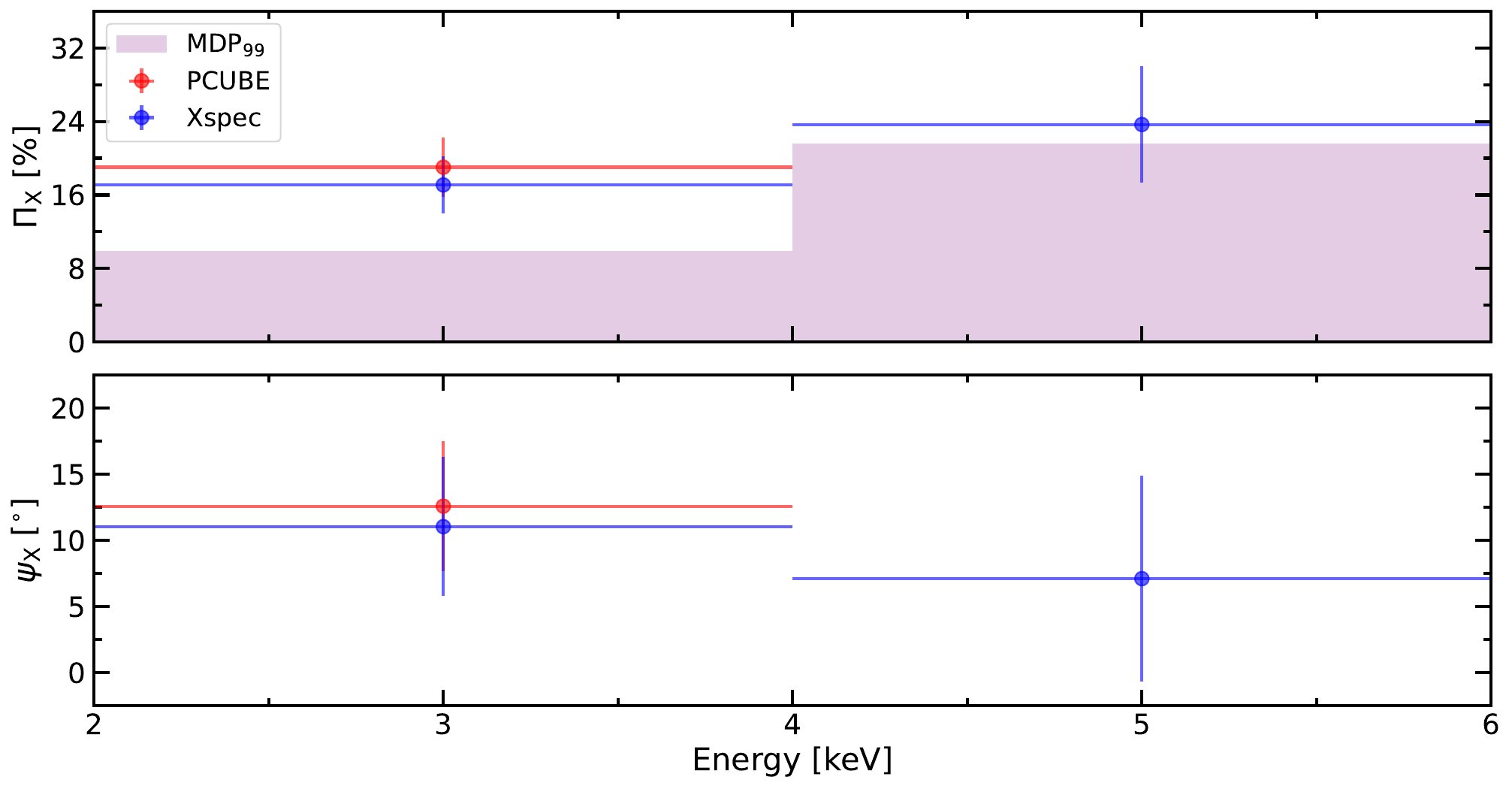} \\
            \makebox{(a)}
    \end{minipage} \\
    \begin{minipage}{0.45\textwidth}
        \centering
        \includegraphics[angle=0, width=\textwidth]{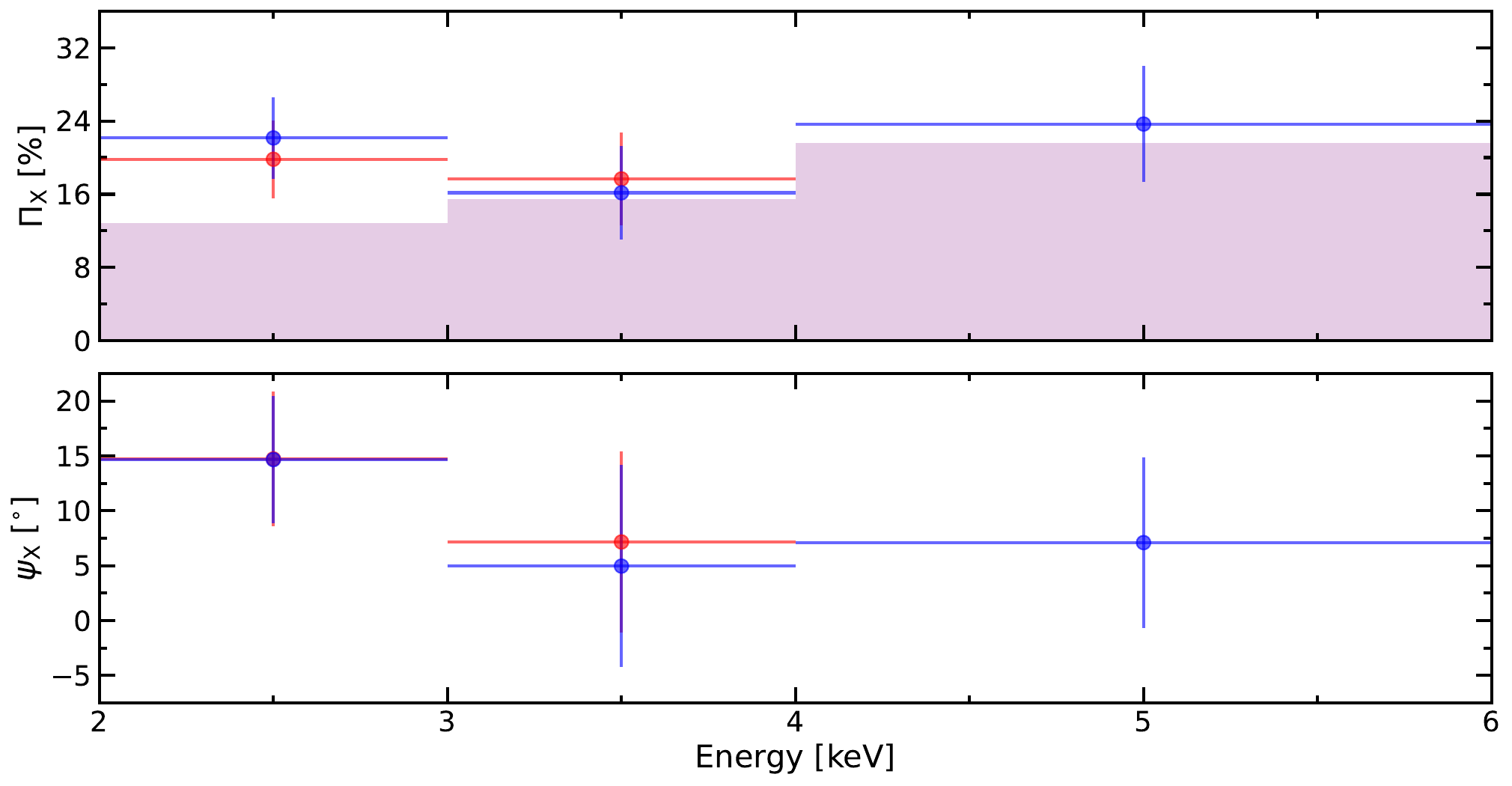} \\
            \makebox{(b)}
    \end{minipage} \\
    \begin{minipage}{0.45\textwidth}
        \centering
        \includegraphics[angle=0, width=\textwidth]{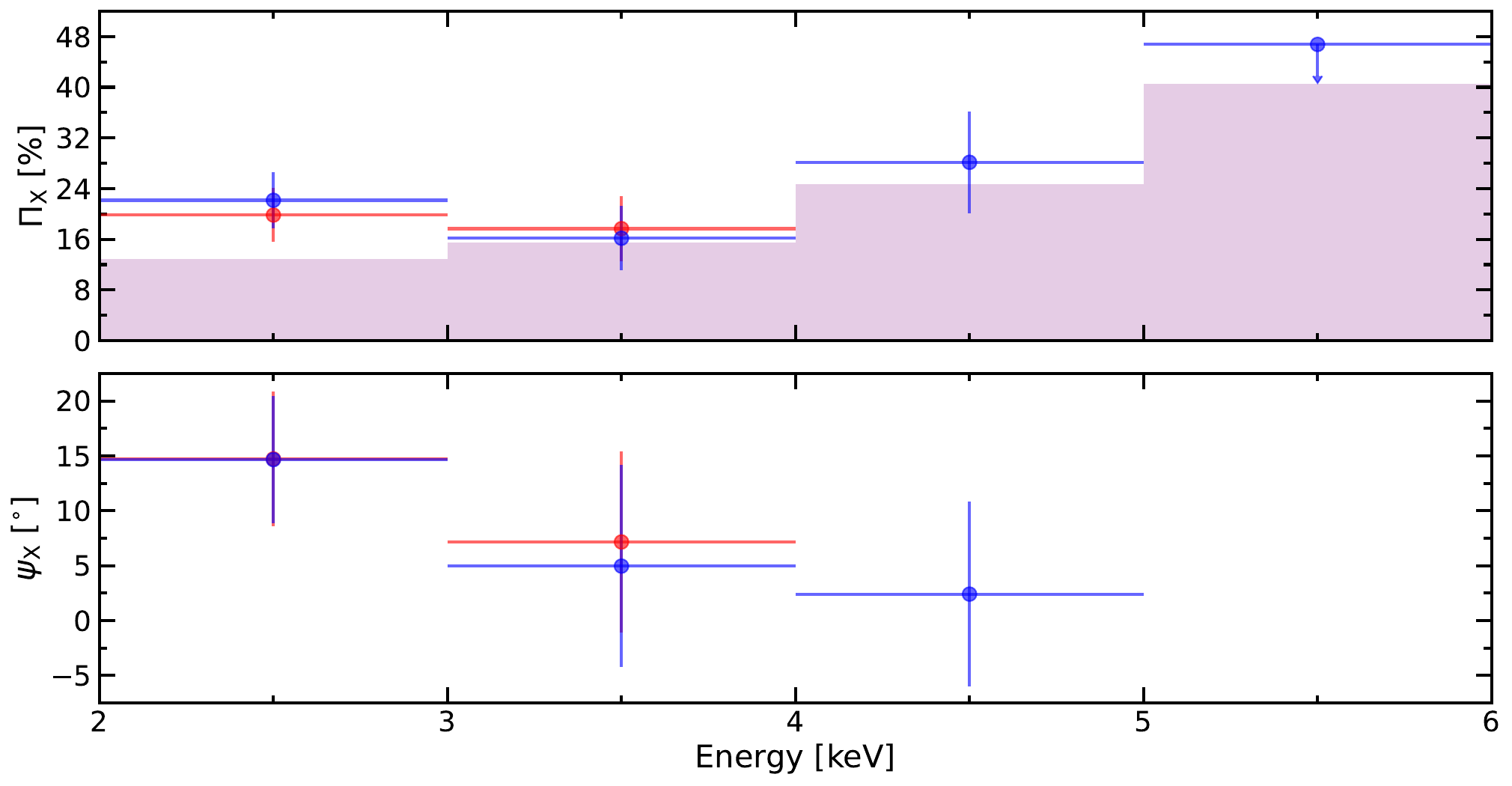} \\
            \makebox{(c)}
    \end{minipage}
    \caption{Results of the energy-resolved polarization analysis covering the entire exposure of the IXPE observation of 1ES 11101--232. Sub-figures (a), (b) and (c) represent the cases that the chosen IXPE energy range is divided into two, three and four energy bins, as described in Section \ref{subsec:ebin}. Indications of the data points and shaded histogram are same with that utilized in Figure \ref{fig:tbin}.}
    \label{fig:ebin}
\end{figure*}

As described in Section \ref{subsec:tbin}, both $\Pi_{\rm X}$ and $\psi_{\rm X}$ of 1ES 1101--232 exhibit a slight variation in the 1-D polarization light curves, accompanied by a spectral variation. Specifically, $\Pi_{\rm X}$ initially decreases and then increases again, coinciding with spectral softening followed by hardening, while $\psi_{\rm X}$ undergoes a slight shift from East to North in the sky projection. To date, the only unambiguous detection of $\psi_{\rm X}$ rotation on timescales of days has been reported in Mrk 421 \citep{2023NatAs...7.1245D,2024A&A...681A..12K}, where such rotation was interpreted as evidence of a helical magnetic field structure in the innermost region of the jet. However, as shown in Figure \ref{fig:custom}, the 2-D contours between $\Pi_{\rm X}$ and $\psi_{\rm X}$ at the 99\% CL for the three custom time intervals are almost overlapping, indicating that no temporal variation was found in the X-ray polarization of 1ES 1101--232 during the IXPE pointing epoch.

\begin{figure*}
    \centering
    \includegraphics[angle=0, scale=0.3]{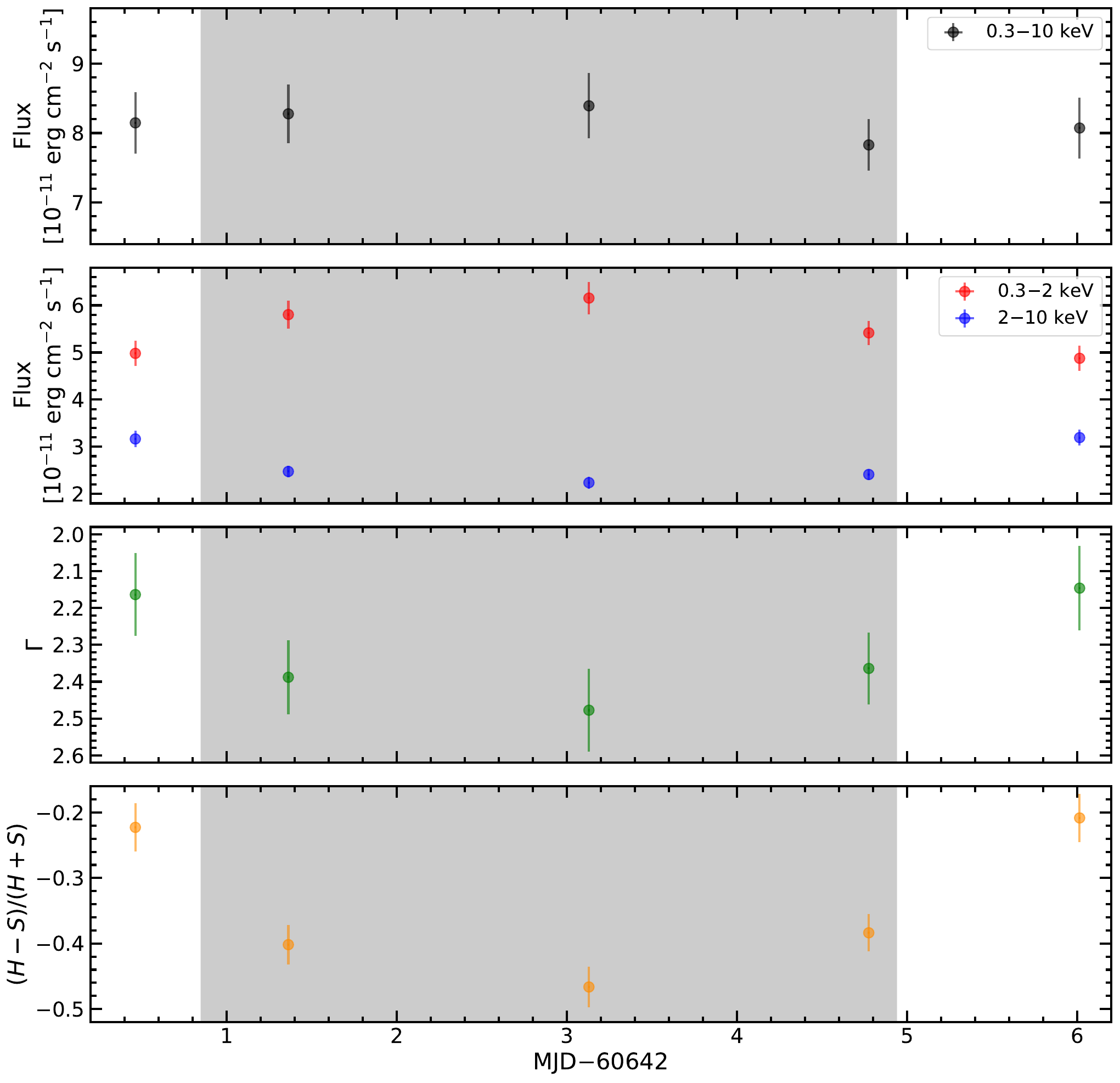}
    \caption{X-ray light curves of 1ES 1101--232, organized with the data of Swift-XRT observations simultaneous or quasi-simultaneous with the IXPE observation. $F_{0.3-10}$ (black points), $F_{0.3-2}$ (red points) and $F_{2-10}$ (blue points), $\Gamma$ (green points), and HR (orange points) are depicted in the first, second, third and fourth panels (from top to bottom), respectively. The gray shaded areas indicate that IXPE pointing epoch.}
    \label{fig:lc}
\end{figure*}

\begin{figure*}
    \centering
    \begin{minipage}{0.3\textwidth}
        \centering
        \includegraphics[angle=0, width=\textwidth]{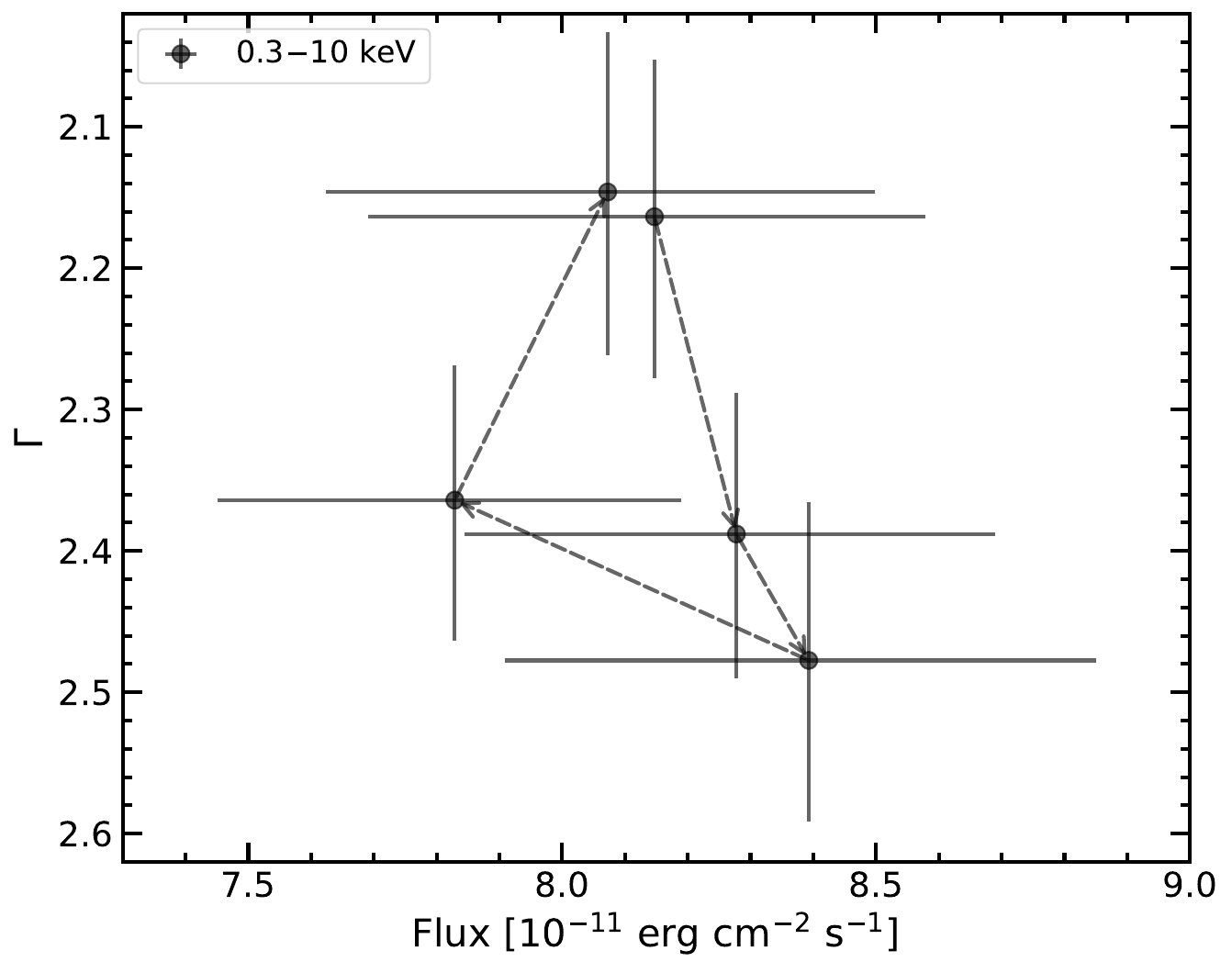} \\
            \makebox{(a)}
    \end{minipage}
    \begin{minipage}{0.3\textwidth}
        \centering
        \includegraphics[angle=0, width=\textwidth]{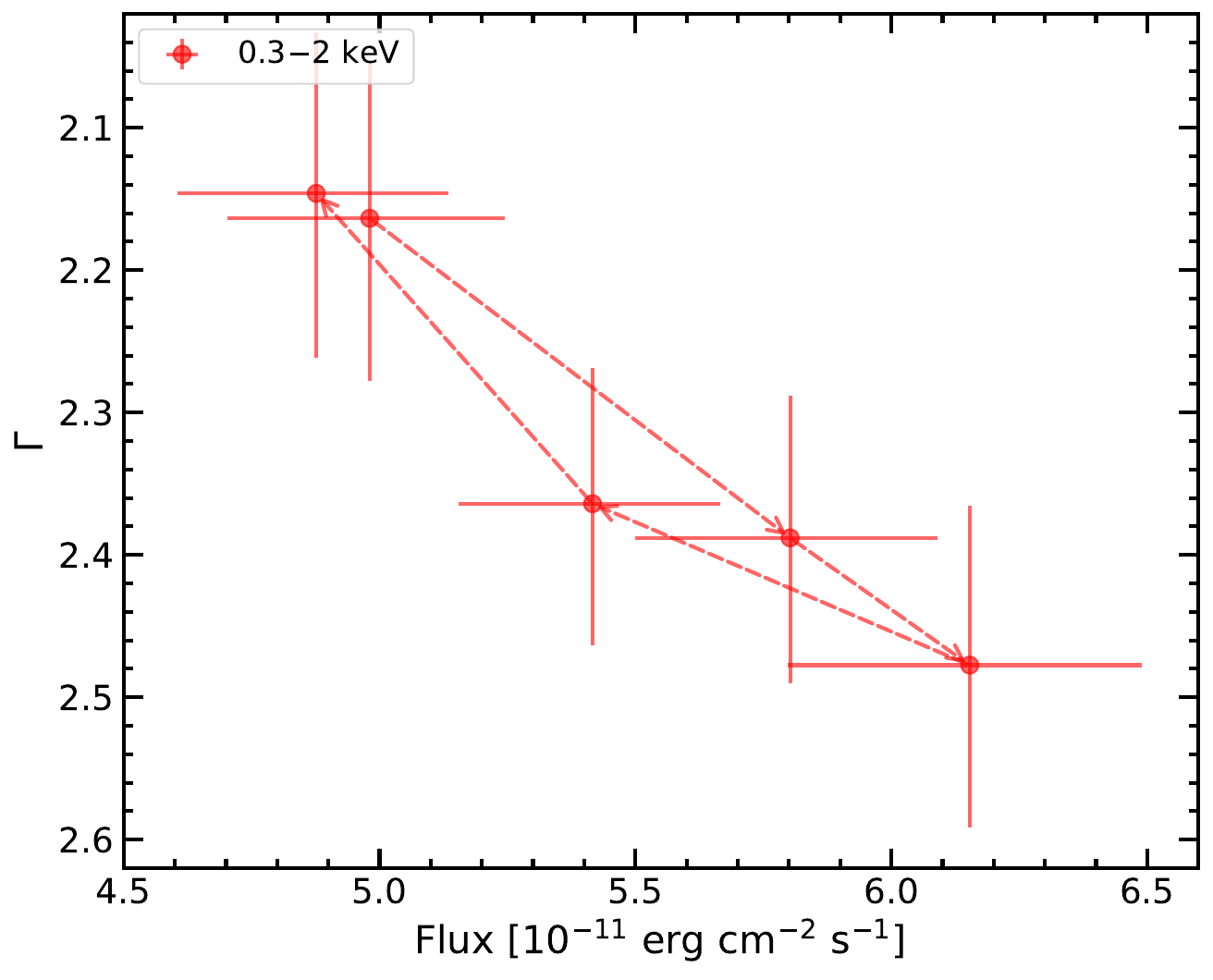} \\
            \makebox{(b)}
    \end{minipage}
    \begin{minipage}{0.3\textwidth}
        \centering
        \includegraphics[angle=0, width=\textwidth]{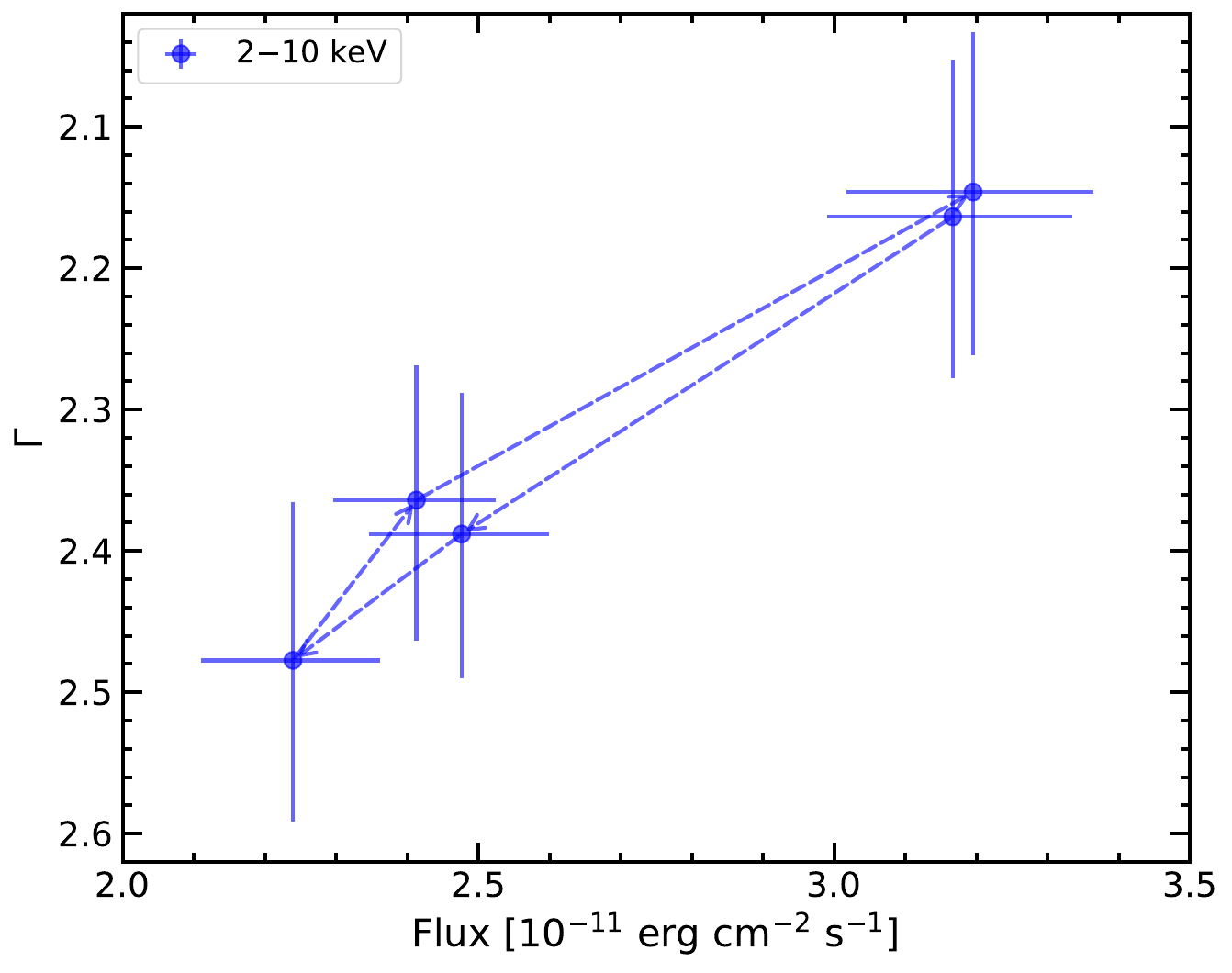} \\
            \makebox{(c)}
    \end{minipage}
    \caption{HIDs of 1ES 1101--232 in the 0.3--10 keV (a), 0.3--2 keV (b) and 2--10 keV (c) bands derived from the Swift-XRT observations.}
    \label{fig:flux-gamma}
\end{figure*}

The Swift-XRT observations of 1ES 1101--232 reveal that the X-ray spectra exhibit a clear soft-to-hard spectral evolution during the IXPE observational period, as illustrated in Figure \ref{fig:lc}. In the 2--10 keV band, the emission demonstrates a {\it harder-when-brighter} behavior (as shown in Figure \ref{fig:flux-gamma}), which is commonly interpreted as indicative of high-energy electron injection via shock (see \citealp{1998A&A...333..452K} and references therein). In contrast, the 0.3--2 keV band displays a {\it softer-when-brighter} trend, while no significant flux variability or correlation between flux and spectral index is detected in the broad 0.3--10 keV band. Notably, clockwise hysteresis loops are observed across all three energy bands, as shown in Figure \ref{fig:flux-gamma}, suggesting that the X-ray emission is dominated by radiation components with shorter cooling times at higher energies, consistent with synchrotron radiation \citep{2004ApJ...601..165F}. Counterclockwise hysteresis patterns, by comparison, would imply comparable electron acceleration and cooling timescales \citep{1998A&A...333..452K}. During the IXPE observation epoch, the observed spectral hardening may be attributed to the re-acceleration or injection of a population of high-energy particles, leading to enhanced flux and a harder spectrum in the 2--10 keV band. The derived photon spectral index of $\Gamma\sim2.45$ indicates that the high-energy tail of the synchrotron emission likely extends into the IXPE energy range, thereby implying a high degree of polarization \citep{2019ApJ...885...76P}. Nevertheless, further multiwavelength observational data are required to validate this interpretation.

\begin{acknowledgments}

This work reports observations obtained with the Imaging X-ray Polarimetry Explorer (IXPE), a joint US (NASA) and Italian (ASI) mission, led by Marshall Space Flight Center (MSFC). The research uses data products provided by the IXPE Science Operations Center (MSFC), using algorithms developed by the IXPE Collaboration, and distributed by the High-Energy Astrophysics Science Archive Research Center (HEASARC).

We sincerely appreciate the valuable suggestions from the referee that improved the manuscript. This work is supported by the National Key R\&D Program of China (grant 2023YFE0117200) and the National Natural Science Foundation of China (grants 12373041, 12373042, 12422306, U1938201, 12022305, 11973050).

\end{acknowledgments}

\clearpage

\bibliography{reference}{}
\bibliographystyle{aasjournalv7}

\clearpage

\begin{table*}
    \begin{center}
    \caption{Observations for 1ES 1101--232 Utilized in This Work}
    \label{tab:obs}
    \begin{tabular}{ccccc}
    \hline
    \hline
    Observatory & OBSID & Date & Net Exposure & Energy Range \\
    & & (YYYY-MM-DD) & (ks) & (keV) \\
    \hline
    IXPE & 03006901 & 2024-11-28 & $\sim191$ & 2--8 \\
    Swift-XRT & 00035013052 & 2024-11-28 & $\sim0.9$ & 0.3--10 \\
    Swift-XRT & 00035013053 & 2024-11-29 & $\sim1.0$ & 0.3--10 \\
    Swift-XRT & 00035013055 & 2024-12-01 & $\sim0.8$ & 0.3--10 \\
    Swift-XRT & 00035013056 & 2024-12-02 & $\sim0.9$ & 0.3--10 \\
    Swift-XRT & 00035013057 & 2024-12-04 & $\sim0.9$ & 0.3--10 \\
    NuSTAR & 61001017002 & 2024-11-29 & $\sim35$ & 3--79 \\
    \hline
    \end{tabular}
    \end{center}
\end{table*}

\begin{table*}
    \begin{center}
    \caption{SpecPol Analysis Results of 1ES 1101--232}
    \label{tab:specpol}
    \begin{tabular}{cccc}
    \hline
    \hline
    Component & Parameter & Unit & Value \\
    \hline
    \multicolumn{4}{c}{Model = \tt{CONSTANT}$\times$\tt{TBABS}$\times$\tt{POWERLAW}} \\
    \hline
    \tt{CONSTANT} & $C_{\rm DU1}$ & & 1.0 (fixed) \\
    & $C_{\rm DU2}$ & & $0.980\pm0.015$ \\
    & $C_{\rm DU3}$ & & $0.980\pm0.014$ \\
    & $C_{\rm XRT}$ & & $1.063\pm0.041$ \\
    & $C_{\rm FPMA}$ & & $1.315\pm0.020$ \\
    & $C_{\rm FPMB}$ & & $1.328^{+0.021}_{-0.020}$ \\
    \tt{TBABS} & $N_{\rm H}$ & $10^{21}~{\rm cm^{-2}}$ & $1.34^{+0.11}_{-0.10}$ \\
    \tt{POWERLAW} & $N_{0}$ & $10^{-2}~{\rm ph~cm^{-2}~s^{-1}~keV^{-1}}$ & $1.62\pm0.03$ \\
    & $\Gamma$ & & $2.45\pm0.01$ \\
    Fit Statistic & $\chi^{2}$/dof & & 881/838 \\
    Flux & $F_{2-8}$ & $10^{-11}~{\rm erg~cm^{-2}~s^{-1}}$ & $1.96\pm0.05$ \\
    \hline
    \multicolumn{4}{c}{Model = \tt{CONSTANT}$\times$\tt{TBABS}$\times$\tt{POLCONST}$\times$\tt{POWERLAW}} \\
    \hline
    \tt{POLCONST} & $\Pi_{\rm X}$$^{*}$ & \% & $17.9\pm2.7$ \\
    & $\psi_{\rm X}$$^{*}$ & $\degr$ & $10.0\pm4.4$ \\
    Fit Statistic & $\chi^{2}$/dof & & 324/343 \\
    \hline
    \end{tabular}
    \end{center}
    \tablenotetext{*}{Polarization parameters are measured in the 2--6 keV band.}
\end{table*}

\begin{table*}
    \begin{center}
    \caption{Swift-XRT Spectral Analysis Results of 1ES 1101--232}
    \label{tab:xrt}
    \begin{tabular}{cccccccc}
    \hline
    \hline
    OBSID & $N_{\rm H}$$^{*}$ & $N_{0}$ & $\Gamma$ & $\chi^{2}$/dof & $F_{0.3-10}$ & $F_{0.3-2}$ & $F_{2-10}$ \\
    \cmidrule(r){6-8}
    & ($10^{21}~{\rm cm^{-2}}$) & ($10^{-2}~{\rm ph~cm^{-2}~s^{-1}~keV^{-1}}$) & & & \multicolumn{3}{c}{($10^{-11}~{\rm erg~cm^{-2}~s^{-1}}$)} \\
    \hline
    00035013052 & 1.34 & $1.57\pm0.10$ & $2.16\pm0.11$ & 26/17 & $8.15^{+0.43}_{-0.46}$ & $4.98^{+0.26}_{-0.28}$ & $3.17^{+0.17}_{-0.18}$ \\
    00035013053 & 1.34 & $1.69\pm0.10$ & $2.39\pm0.10$ & 13/18 & $8.28^{+0.41}_{-0.43}$ & $5.80^{+0.29}_{-0.30}$ & $2.48^{+0.12}_{-0.13}$ \\
    00035013055 & 1.34 & $1.73\pm0.10$ & $2.48\pm0.11$ & 13/15 & $8.39^{+0.46}_{-0.48}$ & $6.15^{+0.34}_{-0.46}$ & $2.34^{+0.12}_{-0.13}$ \\
    00035013056 & 1.34 & $1.59\pm0.09$ & $2.36\pm0.10$ & 45/23 & $7.83^{+0.36}_{-0.38}$ & $5.42^{+0.25}_{-0.26}$ & $2.41^{+0.11}_{-0.12}$ \\
    00035013057 & 1.34 & $1.54\pm0.10$ & $2.15^{+0.12}_{-0.11}$ & 16/16 & $8.07^{+0.43}_{-0.45}$ & $4.88^{+0.26}_{-0.27}$ & $3.20^{+0.17}_{-0.18}$ \\
    \hline
    \end{tabular}
    \end{center}
    \tablenotetext{*}{$N_{\rm H}$ was fixed at the best-fit value obtained from the joint spectra fit discussed in Section \ref{subsec:ixpe}.}
\end{table*}

\clearpage

\end{document}